\documentclass[11pt]{article} 
\usepackage[utf8]{inputenc}
\usepackage{multirow}
\usepackage{amsfonts}
\usepackage{amsmath}
\usepackage{amssymb}
\usepackage{amsthm}
\usepackage{simpler-wick}
\makeatletter
\pgfkeys{simplerwick,wickcolor/.store in=\swick@color,wickcolor=black}
\def\swick@end#1#2{
  \swick@setfalse@#1
  \tikzexternaldisable
  \begin{tikzpicture}[remember picture, baseline=(swick-close#1.base)]
    \node[use as bounding box, inner sep=0pt, outer sep=0pt] (swick-close#1) {$\displaystyle #2$};
  \end{tikzpicture}
  \tikz[remember picture, overlay]
    \draw[\swick@color] ($(swick-open#1.north) + (0, 3pt)$)
          -- ($(swick-open#1.base) + (0, \swick@offset) + #1*(0, \swick@sep)$)
          -- ($(swick-close#1.base) + (0, \swick@offset) + #1*(0, \swick@sep)$)
          -- ($(swick-close#1.north) + (0, 3pt)$);
  \tikzexternalenable}
\makeatother    

\usepackage{simplewick}
\usepackage{float}

\usepackage{caption}
\usepackage{xcolor}
\definecolor{dgreen}{rgb}{0,0.5,0}
\definecolor{dpink}{rgb}{1,0.3,0.3}
\definecolor{darkblue}{rgb}{0,0,0.6}
\definecolor{purple}{rgb}{0.4,.2,0.7}
\usepackage[margin = 2.5cm]{geometry}
\usepackage{graphicx}
\usepackage[hyperfootnotes = false, colorlinks = true, linkcolor = blue, citecolor = purple]{hyperref}
\usepackage{subcaption}

\newcommand{\be}{\begin{equation}}
\newcommand{\ee}{\end{equation}}
\newcommand{\bea}{\begin{eqnarray}}
\newcommand{\eea}{\end{eqnarray}}
\def\la{\label}
\def\nref#1{(\ref{#1})}

\def\tr{\mathrm{tr}}

\def\tl{{\tilde{\ell}}}
\def\tot{\mathsf{tot}}
\def\ca{\mathcal{A}}

	\newcommand{\bes}{\begin{equation} \begin{split} }	
	\newcommand{\ees}{\end{split} \end{equation} }

	\newcommand{\etal}{\textit{et al.}}

	\newcommand{\lp}{\left (}
	\newcommand{\rp}{\right )}
	\newcommand{\lb}{\left [}
	\newcommand{\rb}{\right ]}
	
	\newcommand{\ra}{\rightarrow}

	\newcommand{\pd}{\partial}
	
	\newcommand{\inv}{^{-1}}
	\newcommand{\rt}{\sqrt{2}}
	\newcommand{\hf}{\frac{1}{2}}

	\usepackage{physics}

    \def\nn{\mathcal{N}}
	\def\cj{\mathcal{J}}
	
	\def\cre{\alpha^{\dagger}}

 \def\lt{\mathsf{L}}
 \def\rt{\mathsf{R}}

\def\slt{SL$(2,\mathbb{R})$}
\def\qq{\mathfrak{q}}
\def\rr{\mathfrak{r}}

\begin{document}

\thispagestyle{empty}
\begin{center}
    ~\vspace{5mm}

  {\LARGE \bf {The bulk Hilbert space of double scaled SYK\\}}

   \vspace{0.5in}
     
   {\bf Henry W. Lin$^{1,2}$}

    \vspace{0.5in}

   ~
   \\
   $^1$Institute for Advanced Study,  Princeton, NJ 08540, USA \\
    $^2$Stanford Institute for Theoretical Physics, \\Stanford University, Stanford, CA 94305

    \vspace{0.5in}

    \vspace{0.5in}
    
\end{center}

\vspace{0.5in}

\begin{abstract}
The emergence of the bulk Hilbert space is a mysterious concept in holography. In \cite{Berkooz:2018jqr}, the SYK model was solved in the double scaling limit by summing chord diagrams. Here, we explicitly construct the bulk Hilbert space of double scaled SYK by slicing open these chord diagrams; this Hilbert space resembles that of a lattice field theory where the length of the lattice is dynamical and determined by the chord number. Under a calculable bulk-to-boundary map, states of fixed chord number map to particular entangled 2-sided states with a corresponding size. This bulk reconstruction is well-defined even when quantum gravity effects are important.
Acting on the double scaled Hilbert space is a Type II$_1$ algebra of observables, which includes the Hamiltonian and matter operators.
In the appropriate quantum Schwarzian limit, we also identify the JT gravitational algebra including the physical \slt\, symmetry generators, and obtain explicit representations of the algebra using chord diagram techniques.

\end{abstract}

\vspace{1in}

\pagebreak

\setcounter{tocdepth}{3}
{\hypersetup{linkcolor=black}\tableofcontents}

 \section{Introduction} 
The AdS/CFT correspondence can be summarized in two equations:
\begin{equation}
\begin{split}
\label{adscft1}
 Z_\text{QFT} =   Z_\text{gravity} 
\end{split}
\end{equation}
\begin{equation}
\begin{split}
\label{adscft2}
 \mathcal{H}_\text{QFT} =  \mathcal{H}_\text{gravity}.\end{split}
\end{equation}
The first equation relates the field theory partition function to the gravity partition function. 
The second equation relates the field theory Hilbert space to the Hilbert space of quantum gravity in AdS. %
We don't know how to define precisely the RHS in either \nref{adscft1} or \nref{adscft2}. However, at large $N$ and strong coupling, we expect that the RHS of \nref{adscft1} can be approximated as a path integral over semi-classical geometries. In the same spirit, the RHS of \nref{adscft2} should include black holes states with various excitations of bulk fields. %

In recent years, the SYK model has sparked a flurry of progress \cite{KitaevTalks, Maldacena:2016hyu}. This is in part due to the fact that an analog of equation \nref{adscft1} can be derived from the UV definition of the SYK model \nref{conventions}. 
Integrating out the random couplings and performing a Hubbard–Stratonovich transformation, one can write the SYK partition function as 
\begin{equation}
\begin{split}
\label{}
Z =   \int DG(t,t') \, D\Sigma(t,t')  \, e^{- N I[\Sigma, G] }
\end{split}
\end{equation}
The RHS is analogous to the semi-classical gravity path integral in the sense that the $N$ dependence enters only via the factor of $N$ multiplying the action, akin to the $1/G_N$ in the Einstein-Hilbert action. The ``master field'' variables $G, \Sigma$ have no explicit $N$ dependence, and therefore a saddle point approximation to the integral is reliable at large $N$. 
Furthermore, at low energies, this action reduces to the Schwarzian action of Nearly AdS$_2$ \cite{Almheiri:2014cka, Maldacena:2016upp}, which provides a starting point for the holographic correspondence. 

While the $G,\Sigma$ action has a better understanding of \nref{adscft1}, it seems fair to say that similar progress hasn't been made in understanding \nref{adscft2} in SYK.
Despite the virtues of the $G,\Sigma$ action, one cannot straightforwardly use it to define a bulk Hilbert space. A basic problem is that the fields $G(t,t'), \Sigma(t,t')$  involve two times, so one cannot easily ``cut open'' the path integral to get a Hilbert space. 

In this paper, we will understand \nref{adscft2} in the specialized setting of the SYK model in the strict double scaling limit \cite{Cotler:2016fpe, Berkooz:2018qkz, Berkooz:2018jqr}.
In this limit, \cite{Berkooz:2018jqr} showed that the SYK model can be solved by summing chord diagrams.
A possible reaction to the papers of \cite{Berkooz:2018jqr,Berkooz:2018qkz} is that the chords are ``merely'' a combinatorial gadget that is useful for computations. Here we  advocate {\it against} this view: the chords are precisely the right language to discuss the emergence of the bulk in the double scaled theory.
In particular, the chord diagrams can be naturally ``sliced open'' to construct a 2-sided wormhole Hilbert space. In the simplest case of the pure TFD with no matter perturbations (discussed in Section \ref{chordspure}), this Hilbert space consists of wavefunctions of $\ell$, the length of the wormhole. This $\ell$ variable is actually discrete, with allowed values $ \ell = (2 q^2 /N) n$ where $n$ is a non-negative integer. However in the quantum Schwarzian limit, this discreteness is negligible and one recovers the Hilbert space of pure JT gravity.

In Section \ref{matter}, we consider more general states obtained by perturbing the TFD with matter operators. 
The chord construction gives a bulk Hilbert space which resembles that of a lattice field theory in 1+1 dimensions where the overall length $\ell$ of the lattice (the chord number) is allowed to fluctuate. 
What's more, besides simply identifying the bulk Hilbert space in \nref{adscft2}, we would like to know concretely how states in the bulk Hilbert space are mapped to states in the boundary Hilbert space. 
We present in Section \ref{bulkBd} a calculable bulk-to-boundary map that reconstructs states of the wormhole with various particle excitations in the boundary theory. An interesting property of this reconstruction is that the bulk chord number is mapped to the size of the operator used to create the 2-sided boundary state. The discreteness of the bulk length can thus be traced to the boundary fact that operator size is an integer.

Our construction is related to the recent discussion of emergent large $N$ algebras \cite{Leutheusser:2021qhd, Leutheusser:2021frk, Witten:2021unn, Chandrasekaran:2022cip}.
In Section \ref{bulkBd}, we define a type II$_1$ algebra of observables $\hat{\mathcal{A}}$ in the strict double scaling limit. Given our explicit bulk-to-boundary map, this algebra can be thought of as acting either on the boundary or the bulk Hilbert space. A key difference relative to \cite{Leutheusser:2021qhd, Leutheusser:2021frk} is that the Hamiltonian is part of the double scaled algebra. The explanation of this is simple: in SYK the Hamiltonian scales $\sim N/q^2$. For regular SYK in the $N \to \infty$ limit, the Hamiltonian is divergent and not part of the large $N$ algebra as in \cite{Leutheusser:2021qhd,Leutheusser:2021frk}. But in the double scaling limit, we also take $q\to \infty$ holding $N/q^2$ fixed, so Hamiltonian remains finite and can therefore be included in the double scaled algebra. This is closely related to the statement that in the double scaled limit, quantum gravity effects are still important.

An elementary feature of holography is the identification of bulk symmetries with boundary symmetries. In JT gravity, the physical \slt\, symmetries that move matter around in nearly AdS$_2$ are somewhat subtle \cite{Lin:2019qwu}. In particular, they do not commute with the Hamiltonian. Instead, the Hamiltonian is part of a larger ``gravitational algebra'' \cite{Harlow:2021dfp} that includes \slt\, as a subalgebra \cite{Lin:2019qwu}. In Sections \ref{matter} and \ref{bulkBd}, we identify a subalgebra $\mathcal{G} \subset \mathcal{A}$ that satisfies the gravitational algebra in the appropriate quantum Schwarzian limit. For wormholes with $m$ particles in the interior, we obtain concrete representations of the gravitational algebra, where $\mathcal{G}$ acts on wavefunctions of multiple lengths $\ell_0, \ell_1, \cdots, \ell_m$, generalizing the $m=0$ case of Liouville quantum mechanics \cite{Bagrets:2016cdf, Harlow:2018tqv}. Away from this limit, we conjecture that $\mathcal{G}$ satisfies a suitable $\qq$-deformation of the JT gravitational algebra.

The sections in this paper are roughly ordered by increasing generality/abstraction. For example, Section \ref{sec:dsalg} can be read as a summary (in a somewhat more abstract language) of the concrete calculations presented in previous sections. %
Hopefully the reader will find this organization pedagogical.  %

\subsection{Conventions}
We adopt the following definitions for the SYK model:%
\begin{equation}
\begin{split}
\label{conventions}
   &\left\{\psi_{i}, \psi_{j}\right\}=2 \delta_{i j},\\
&H =i^{q / 2} \!\!\!\!\!\!\!\!\sum_{1 \leq i_{1} <\cdots < i_{q} \leq N} \!\!\!\!\!\!J_{i_{1} \ldots i_{q}} \psi_{i_{1}} \cdots \psi_{i_{q}}, \quad
\left\langle J_{i_{1} \ldots i_{q}}^{2}\right\rangle
=\frac{\cj^2}{\lambda {N \choose q} },\\
&\lambda = 2q^2/N,  \quad \qq = \exp(-\lambda).
\end{split}
\end{equation}
With these conventions, $\tr H^2 = \cj^2/ \lambda $, where $\tr$ is the normalized trace $\tr \mathbf{1} = 1$ . %
The {\it double scaled} SYK model refers to the limit where $q \to \infty, \, N \to \infty, \, \lambda = 2q^2/N $ held fixed. The {\it triple scaling} limit is a further limit where $\lambda \ll 1$ and energies $E/\cj \ll 1$. In this limit, the theory is governed by the Schwarzian action with coupling $C = N/(4q^2 \cj)$.   %
 If we study the triple scaled theory in a regime $C/\beta \ll 1$, the quantum fluctuations of the Schwarzian mode are large. We henceforth set $\cj = 1$.
We will also adopt the shorthand $I = \{ 1 \le i_1 <  \cdots <  i_q \le N\} $ to denote sets of indices, and denote products of fermions via
\begin{equation}
\begin{split}
\label{}
  \Psi^s_{I} =\psi_{i_{1}} \cdots \psi_{i_{s}}.
\end{split}
\end{equation}
With this notation, $H = i^{q/2} \sum_I J_I \Psi^q_I$.
For the convenience of the reader, we note that Berkooz \etal\, \cite{Berkooz:2018jqr} uses the following conventions: $ q_\text{Berkooz} = \qq$, $\lambda_\text{Berkooz} = \lambda$, $p_\text{Berkooz}=q$ and $\cj_\text{Berkooz}^2 = \cj^2 /\lambda$. Thus our formulas involving the Hamiltonian will differ by a factor of $\sqrt{\lambda}$ from Berkooz \etal\,
The normalization of our Hamiltonian \nref{conventions} is natural from the point of view of studying the SYK model at finite $q$. (For example, the max energy scales linearly with $N$ in this conventions). In the double scaling limit, it is admittedly more natural to set $\tr H^2 = 1$ as in \cite{Berkooz:2018jqr}; nevertheless, we will stick with \nref{conventions} out of familiarity.

\section{Chords and the 2-sided Hilbert space \la{chordspure} }

\subsection{Review of double scaled SYK}
Here we review the chord diagram technique in double-scaled SYK \cite{Cotler:2016fpe, Berkooz:2018jqr, Berkooz:2018qkz}. Let us consider the computation of the partition function $Z = \tr (e^{-\beta H})$, which is carried out in the following steps:
\begin{enumerate}
	\item Expand $e^{-\beta H} = \sum_k \beta^k  H^k/k!$ and focus on evaluating the sum term by term.
	\item Perform the disorder average. This amounts to Wick-contracting the $J$'s present in each factor of $H$.
	\item For each Wick contraction, evaluate the remaining factor $\tr (\Psi_{I_1}^q \Psi_{I_2}^q \cdots \Psi_{I_1}^q \cdots )$.  
	 To evaluate this trace, we need to anti-commute like fermions next to each other and then annihilate them using $\psi_i^2 = 1$.  If we anti-commute two sets of fermions $\Psi_{I_1}^q$ and $\Psi_{I_2}^q$ past each other, we get a possible minus sign $(-1)^f$ depending on whether $f = |I_1 \cap I_2|$ is even or odd. In the double-scaling limit, we can replace this phase by the average value\footnote{The number $f$ of fermions in common is Poisson distributed with mean $q^2/N = \lambda/2$. The expectation value of $(-1)^f$ in this Poisson distribution is $e^{-\lambda}$. } of the phase $\qq = \ev{ (-1)^f } = e^{-\lambda}$.
	 \item Sum over all possible Wick contractions and also over $k$. This is the subject of the next subsection. %
\end{enumerate}
The idea is that step 2 and 3 can be summarized by considering the ``chord diagram'' associated to the Wick contraction. This is basically the diagram we are taught in field theory textbooks to draw when performing Wick contractions. More specifically, one draws a chord for each pair in the Wick contraction in such a way that any two chords intersect at most once; an example is drawn in Figure \ref{hilbert}b for $\tr H^k$ where $k = 20$. 
To account for the sign in step 3, we write down a factor of $\qq$ for each intersection of chords (an ``interaction vertex'').

\begin{figure}[t]
    \begin{center}
	\vspace{-3cm}
    \includegraphics[width=\columnwidth]{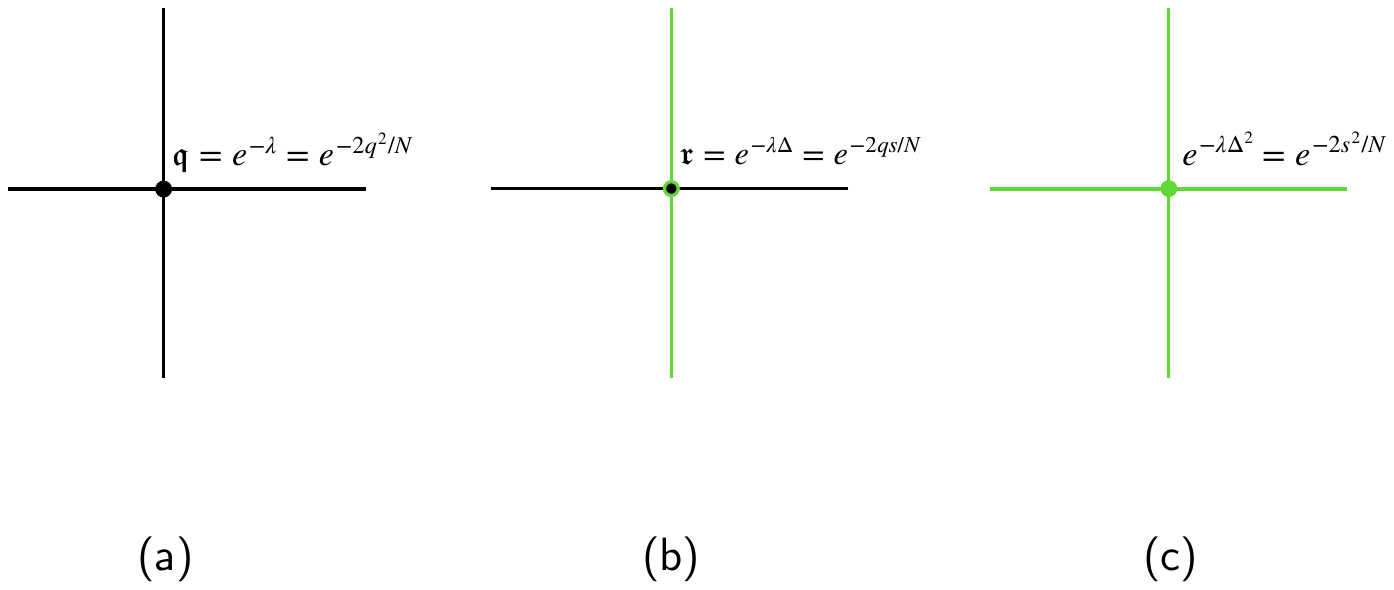}
    \vspace{-3cm}
    \end{center}
    \caption{(a) intersection of two $H$ chords. This gives a factor of $\qq$. (b) Intersection of $H$ (black) and $M$ (green) chords. (c) Intersection of two $M$ chords.}
    \label{feynman}
\end{figure}

Following \cite{Berkooz:2018jqr, Berkooz:2020xne}, one can also use the same technique to compute correlation functions of operators of certain ``matter'' operators, which are monomials in $\psi_i$ of degree $s$ with random coefficients:
\begin{equation}
\begin{split}
\label{matterop}
  M_s %
  = i^{s/2} \sum_I K_{I} \Psi^s_{I}.
\end{split}
\end{equation}
Here $K_I$ is a collection of Gaussian random variables, drawn independently from $J_I$. $K_I$ is normalized so that $\tr M_s^2 = 1$. In the double-scaling limit, one takes $s \to \infty$ holding $\Delta = s/q$ fixed. 
To compute matter correlators $\tr (e^{-\tau_1 H} M_s e^{-\tau_2 H} M_s  \cdots)$ one follows a similar algorithm. The only complication is that one must introduce chords of different types, indicated in our figures as black $(H)$ or green $(M_s)$. Intersections of different chord types get different factors, see Figure \ref{feynman}.
We can consider multiple species of $M_s$, possibly with different sizes $\{M_s, M_{s'}, M_{s''}, \cdots \}$. For each species, the $K_I$ are drawn independently. The interaction vertex for two chords (associated with operators of size $s_1$ and $s_2$)	 crossing is given by $\exp(-2 s_1 s_2 / N)$.

\subsection{The 2-sided chord Hilbert space \la{sec:pure} }
In this subsection, we explain how the bulk Hilbert space emerges in the simplest case of the ``empty wormhole'', where we simply consider the TFD state at various temperatures/Lorentzian times without inserting any matter operators $M_s$. In subsequent sections, we will generalize to the case with matter. Although we could describe the construction without reference to JT gravity, it is useful to have JT gravity in mind.

\begin{figure}[t]
    \begin{center}
    \includegraphics[width=\columnwidth, trim={0, 3.5cm, 0, 4.5cm}, clip]{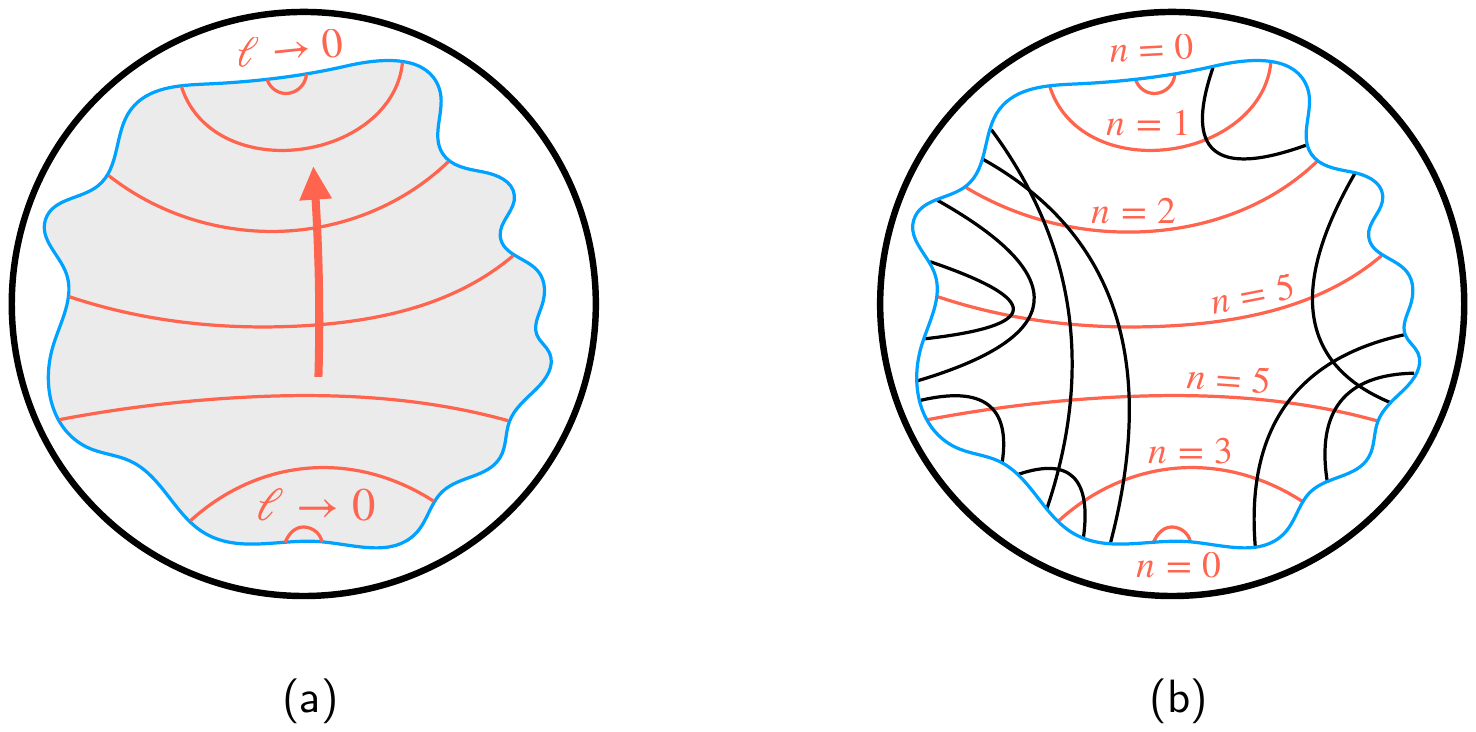}
    \end{center}
    \caption{(a) The Hilbert space of JT gravity with no matter. The dynamical degree of freedom is the length $\ell$ between the two sides (pink curves). The disk partition function is obtained by imposing $\ell=0$ boundary conditions in the Euclidean past and future. (b) The Hilbert space of double scaled SYK with no matter insertions. The dynamical variable is $n$, the number of chords. In the microscopic description $ q n$ is just the size of the density matrix. We have drawn the chords so that for any red curve, the chords which intersect that curve cross do not cross in the Euclidean past.  }
    \label{hilbert}
\end{figure}

To this end, recall that in pure JT gravity (e.g. JT gravity with no matter fields), the classical phase space is just the geodesic length $\ell$ of the 2-sided wormhole and its conjugate momentum \cite{Harlow:2018tqv}. The bulk Hilbert space therefore consists of wavefunctions $\psi(\ell)$. Furthermore, 
in a recent paper (Appendix H of \cite{LongPaper}, see also \cite{Blommaert:2018oro, Blommaert:2018iqz} for a gauge theory perspective) it was pointed out that this 2-sided wormhole Hilbert space in fact arises in the computation of the 1-sided thermal partition function: 
 \begin{equation}
\begin{split}
\label{jt1}
  \tr \lp e^{-\beta H} \rp  \propto \bra{\ell = 0} e^{-\beta H_\text{Liouville}} \ket{\ell = 0}.
\end{split}
\end{equation}
Here $\ell$ is the length between the two sides of the wormhole and $\ket{\ell=0}$ is a position eigenstate $\psi(\ell) = \delta(\ell)$. (More precisely, we should work with the renormalized length $\tl$, replace $\ket{\ell=0}$ with $\psi(\tl) = \delta(\tl-\tl_c)$, and take $\ell_c \to -\infty$.) The idea is that the thermal circle can be viewed as a left and a right boundary, joined at a point in the Euclidean past and in the future, see Figure \ref{hilbert}. 
When the boundaries join, the length between them goes to zero, which is enforced by projecting onto $\ket{\ell = 0}$ and $\bra{\ell = 0}$.  %
We emphasize that the Hilbert space appearing on the LHS of \nref{jt1} is {\it different} than the one appearing on the RHS. The Hamiltonian that appears on the LHS is interpreted as the 1-sided Hamiltonian of the dual quantum mechanical theory (in pure JT gravity, it is a random matrix \cite{Saad:2019lba}), whereas $H_\text{Liouville}$ acts on the 2-sided wormhole Hilbert space as \cite{Bagrets:2016cdf, Harlow:2018tqv} 
\begin{equation}
\begin{split}
\label{jt2}
  H_\text{Liouville} = \frac{1}{2C}\lp - \pd_\tl^2 + e^{-\tl}\rp.
\end{split}
\end{equation}
Notice that by cutting open the diagram in Figure \nref{hilbert}, we can also write %
\begin{equation}
\begin{split}
\label{tfd}
e^{-\beta H/2} \ket{\Omega}  \, \cong \, e^{-\beta H_\text{Liouville}/2 }\ket{\ell = 0}
\end{split}
\end{equation}
We use $\cong$ instead of $=$ because the two states live in different Hilbert spaces. On the LHS of \nref{tfd}, we have a state in the 2-sided microscopic Hilbert space. $\ket{\Omega}$ is a maximally entangled state and $e^{-\beta H} \ket{\Omega} $ is the thermofield double.
On the RHS, the state lives in the Hilbert space of the wormhole. Of course, the meaning of the equation is that the two states are mapped to each other under the bulk-to-boundary holographic map. %

The point of this subsection is to explain the precise analogs of equations \nref{jt1}, \nref{jt2}, and \nref{tfd} in double scaled SYK.
The formulas that appear in this subsection are not new for the most part (they were presented in \cite{Berkooz:2018jqr}) but we will present them here with a clear bulk interpretation, which generalizes to the novel cases with matter.
  
Following the previous paragraph on JT gravity, we also consider the 1-sided thermal partition function. This is given by a sum over chord diagrams. The idea is that we can focus on a particular chord diagram in the sum, and slice it open along a red curve that goes from the left to right boundary, see Figure \ref{hilbert}b. On this curve, one assigns a state in a 2-sided Hilbert space $\ket{n}$, which is the number of open chords\footnote{On a given time slice $t$, let $C_t$ be chords that intersect the time slice. The {\it open chords} are the subset of $C_t$ which have not intersected any of the other chords in $C_t$ in the Euclidean past. It is always possible to choose a slicing such that all chords in $C_t$ are open chords. Note that this definition of open chords is not time-reversal symmetric. We have defined ket states; to define bra states, one can take the time reversal of this definition. This is compatible with our inner product, see Figure \nref{inner}.}
 intersecting this slice. 
As depicted in Figure \ref{hilbert}b, any chord diagram that arises from the thermal circle computation can be sliced in such a way so that at some time in the Euclidean past the two-sided state has no open chords $\ket{0}$, and similarly at some future time the state is again $\bra{0}$. Therefore, we can write
 \begin{equation}
\begin{split}
\label{partition}
  \tr \lb  \exp(-\beta H) \rb = \bra{0} e^{-\beta T} \ket{0},\\
\end{split}
\end{equation}
This equation is the analog of \nref{jt1}, where the (integer) chord number plays the role of the length basis. The analog of $H_\text{Liouville}$ is the matrix
$T$ that is determined by the combinatorics of chord diagrams. Suppose we consider a 2-sided state with fixed chord number $\ket{n}$. We can act on this state with $H_\lt$ or $H_\rt$. On the thermofield double, $H_\lt = H_\rt = H$ so we do not need to distinguish them, although in later sections when we will. The idea is that each time we insert some factor of $H$, an open chord can be created or annihilated. There is only one way to create an open chord, but one can choose any of the $n$ open chords to annihilate:
\begin{figure}[H]
    \begin{center}
    \includegraphics[width = \columnwidth, trim={0, 3.5cm, 0, 6cm}, clip]{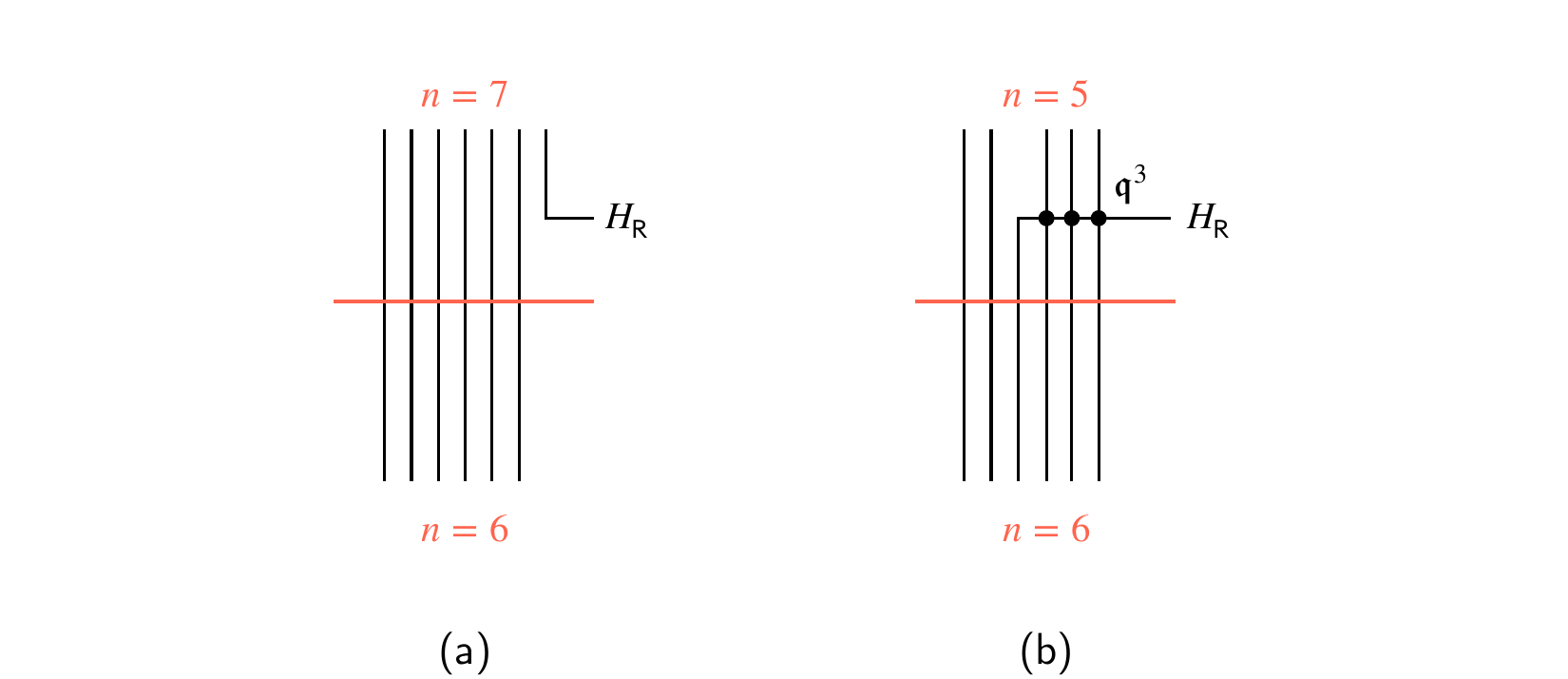}
    \end{center}
    \caption{Two processes that can happen when we act with $H_\rt$. In (a) an extra chord is added. In (b) a chord is removed. It crosses 3 chords giving a factor of $\qq^3$. We could have also consider the action of $H_\lt$. The same processes happen but we would draw the insertion on the left. }
    \label{chords-t}
\end{figure}

\noindent Each choice of a chord to be annihilated gives a different power of $\qq$, corresponding to the number of intersections with chords to the right ($H_\rt$) or left $(H_\lt)$. This gives %
\def\T{\mathrm{T}}
\begin{equation}
\begin{split}
\label{transferBerkooz}
    T  =  \frac{1}{\sqrt{\lambda }} \lp \cre  + \alpha W \rp \\ 
     W_n  = \qq^0 + \qq^1 + \cdots + \qq^{n-1} = \frac{1-\qq^{n} }{1-\qq},\\  \alpha = a \sqrt{1 \over n}, \quad \cre = \sqrt{1\over n}a^\dagger
\end{split}
\end{equation}
The factor of $1/\sqrt{\lambda}$ is just an overall normalization that comes from our convention  $\tr H^2 = \lambda\inv$. $W$ is an operator that is diagonal in the $\ket{n}$ basis with eigenvalues given above, and $a,a^\dagger$ are the raising and lowering operators $[a,a^\dagger ] = 1, [n,a^\dagger] = a^\dagger, [n,a] = -a$.

So far we have defined a bulk vector space spanned by states with definite chord numbers $\ket{n}$. To upgrade this to a Hilbert space, we must specify the inner product. 
The chord diagrams suggest a natural definition. 
Let us insert a complete set of states in the computation of \nref{partition}:
\begin{equation}
\begin{split}
\label{eq:decomp}
\bra{0} T^{a+b} \ket{0} &=   \sum_m (T^a)_{0,m} (T^b)_{m,0} %
= g^{mn}  (T^a)_{m,0} (T^b)_{n,0}. 
\end{split}
\end{equation}
Here $g$ has the interpretation as the inner product $g^{mn} = \braket{m}{n} $. This expression is in 1-to-1 correspondence with a decomposition of the chord diagrams illustrated in Figure \ref{inner} below:
\begin{figure}[H]
    \begin{center}
    \includegraphics[width = \columnwidth, trim={0, 5.5cm, 0, 5.5cm}, clip]{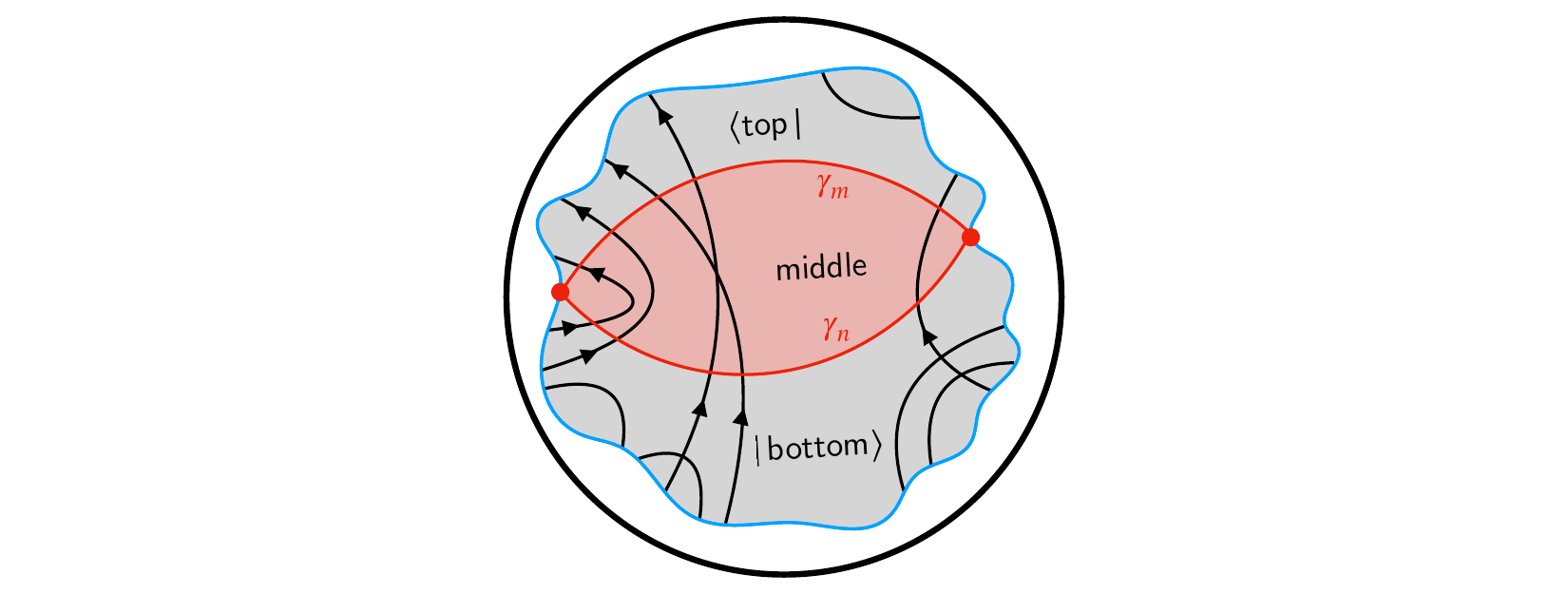}
    \end{center}
    \caption{Interpretation of the chord diagram as an overlap of a bra $\bra{\text{top}}$ and ket $\ket{\text{bottom}}$ defined in the bulk Hilbert space. The middle region ({\color{dpink} pink}) defines an inner product between states. It is defined as a 1-way region: all chords entering through $\gamma_n$ must exit through $\gamma_m$. }
    \label{inner}
\end{figure}

\noindent Any chord diagram decomposes into three regions: top, middle, bottom. The top region defines the bra state. The bottom region defines a ket state. The middle region describes the overlap $g^{mn} = \braket{m}{n}$. The middle region is defined by requiring that all chords cross $\gamma_m$ and $\gamma_n$ exactly once. Using this decomposition, we see that the overlap $\braket{m}{n}$ can be computed by summing all diagrams where $n$ lines enter the bottom of the middle region and $m$ lines exit the top\footnote{These rules were discussed in \cite{Pluma:2019pnc} and \cite{Berkooz:2020xne}.}. Thus if we know $g^{mn}$, we could also evaluate $\tr \lp H^{a+b} \rp$ by summing over the top, middle, and bottom regions, which gives the factors $(T^a)_{m,0}$, $g^{mn}$, and $(T^b)_{n,0}$, respectively in \nref{eq:decomp}.

It is not hard to directly compute $g_{mn}$; the eager reader should jump to \nref{eq:smat_sol}. 
Here we compute it recursively, foreshadowing the case with matter.
To do so, imagine removing one of the chords in the middle region, say the first chord that intersects $\gamma_n$ on the left. This chord can intersect any number from 0 to $n-1$ of its neighbors before passing through $\gamma_m$. After deleting the chord, we are left with the inner product involving one less chord $\braket{m-1}{n-1}$:
\begin{equation}
\begin{split}
\label{eq:smat}
\braket{m}{n} =  W_n \braket{m-1}{n-1}. %
\end{split}
\end{equation}
By iterating this relation, we reduce to evaluating either $\braket{0}{n}$ or $\braket{m}{0}$:%
\begin{equation}
\begin{split}
\label{eq:smat_bd}
\braket{0}{n} = \braket{m}{0} = 0  \text{ for } m,n>0, \quad 
\braket{0}{0} = 1. 
\end{split}
\end{equation}
These vanish for $m, n > 0$ since the middle region is a 1-way region. This implies that states with different chord numbers are orthogonal, and so we write\footnote{I thank Cynthia Yan for pointing out a typo in a previous version of the draft.}: %
\begin{equation}
\begin{split}
\label{eq:smat_sol}
  \braket{n}{m} = S_{n} \delta_{n,m}, \quad 
 S_{n} = S_{n-1} W_n,  \quad S_0 = 1\\
  \quad  S = \prod_{i=1}^{n} {1-\qq^n \over 1- \qq }   = \frac{(\qq, \qq)_n}{(1-\qq)^n}.
\end{split}
\end{equation}
In the second line, we solved the recursion relation. Here  $(a,\qq)_n =\texttt{QPochhammer[}a,\qq,n\texttt{]} $ in Mathematica.
Given this inner product, it is natural to compute overlaps\footnote{To clarify the notation, let $(\cdot , \cdot )$ be the inner product of two (ket) vectors. 
Then $\bra{n}T\ket{m} = \lp \ket{n}, T \ket{m}\rp  = \sum_p \lp \ket{n}, \ket{p}\rp   T_{pm} = \ev{n|p}T_{pm} $. } in canonically normalized states
\begin{equation}
\begin{split}
\label{}
\frac{  \bra{n} T \ket{m}}{\sqrt{\braket{n}} \sqrt{\braket{m}} } = \sqrt{\frac{\braket{n}{n}}{\braket{m}{m}}} T_{nm} \doteq -H_{nm}
\end{split}
\end{equation}
In the last line, we have defined\footnote{The spectrum of $H$ is symmetric under $H \to -H$. Therefore there is a sign ambiguity $H = \pm STS$. We choose the minus sign so that the minimum energy state has small momentum $k\to 0$. } $H = -g^{1/2} Tg^{-1/2}$. 
Using \nref{eq:smat}, this yields a manifestly Hermitian matrix\footnote{Here $W$ is defined as the operator $W \ket{n} = W_n \ket{n}$. }:
\begin{equation}
\begin{split}
\label{}
-H \sqrt{ \lambda }=    \alpha\sqrt{W} + \sqrt{W} \cre 
\end{split}
\end{equation}
This Hamiltonian is the appropriate generalization of the Liouville Hamiltonian \nref{jt2}.
To make contact with JT gravity, we can define a rescaled chord number %
\begin{equation}
\begin{split}
\label{}
  \ell = -n \log \qq = \lambda  n .\\
\end{split}
\end{equation}
\begin{equation}
\begin{split}
\label{largeqHam}
  H &= - {1 \over \sqrt{\lambda (1-\qq) }} \lb  e^{i \lambda k} \sqrt{1- e^{-\ell }} + \sqrt{1- e^{-\ell} }  e^{-i \lambda k  }  \rb%
\end{split}
\end{equation}
In general $k$ is the conjugate momentum to $\ell$ (the generator of $\ell$ translations). In the approximation when $\ell$ is continuous, we can write $k = - i \pd_\ell$.
Note that the Hamiltonian is periodic in the momentum $k \to k + 2\pi /\lambda $. This is expected since its conjugate variable $\ell$ is discretized to take values of $\ell = \lambda n$. From \nref{largeqHam}, we can see immediately that the eigenstates are scattering states with momentum $\pm k$. The eigenvalues of such states are determined by the Hamiltonian at $\ell \to \infty$, which is 
\begin{equation}
\begin{split}
\label{eigenH}
  E(k) =  -\frac{2  \cos \lambda k}{\sqrt{\lambda(1-\qq) } } .
\end{split}
\end{equation}
We can now consider the limit where $\qq \to 1$ (the standard large $q$ SYK model) while focusing on low energies, where we expect a quantum Schwarzian description. This is known as the triple scaling limit, which in this context means
\begin{equation}
\begin{split}
\label{eq:tripleS}
 \lambda \to 0, \quad \ell \to \infty, \quad  e^{-\ell}/\lambda^2 =  e^{-\tl} =\text{fixed} .
\end{split}
\end{equation}
The variable $\tl$ is known as the ``renormalized length'' in the JT literature.
This gives us a precise match with the gravitational Hamiltonian \nref{jt2} including the correct value of the Schwarzian coupling $C$:
\begin{equation}
\begin{split}
\label{eq:tripSHam}
H - E_0 = \frac{1}{2C}\lp k^2 + e^{-\tl}\rp , \quad 2C = 1/\lambda, \quad E_0 = - 2/\lambda.
\end{split}
\end{equation}
The interpretation of the chord number as a quantized length was already suggested in \cite{Berkooz:2018qkz}, however, at the time the 2-sided interpretation of the Liouville quantum mechanics was unclear. 
Here we have derived this correspondence by identifying the chord Hilbert space with the 2-sided bulk Hilbert space. 
As in JT gravity, we should distinguish $H$, the microscopic Hamiltonian defined in \nref{conventions} and the expression for $H$ given by \nref{largeqHam}. They are identified under the bulk-to-boundary map, but the latter expression for $H$ only holds on the subspace of states with no matter insertions. 

As a technical aside, let us note that we do not need to solve the recursion relation \nref{eq:smat} in order to derive $H$ in the triple scaling limit. Instead, we can simply take $S_{n+1} \approx S_n/ \lambda$. Then expanding in powers of $\lambda$ gives %
\begin{equation}
\begin{split}
\label{eq:directH}
 - H \approx \frac{1}{\lambda} \lb e^{\lambda \pd_\ell} + e^{-\lambda \pd_\ell} (1- e^{-\ell}) \rb \approx \frac{2}{\lambda} + \lambda \lp \pd_\ell^2 - \lambda^{-2} e^{-\ell} - \pd_\ell e^{-\ell} \rp 
  \end{split}
\end{equation}
In the triple scaling limit \nref{eq:tripleS}, the last term in \nref{eq:directH} is negligible, so we recover \nref{eq:tripSHam}. In the next section, we discuss the case with matter in the wormhole. The generalization of the recursion relation \nref{eq:smat} is more complicated, but we can still take the triple scaling limit rather easily along these lines.

\section{The wormhole Hilbert space with matter \la{matter} }
 \subsection{Chords and a particle}
 
  \begin{figure}[t]
     \begin{center}
    \includegraphics[width = \columnwidth, trim={0, 3.5cm, 0, 3.5cm}, clip]{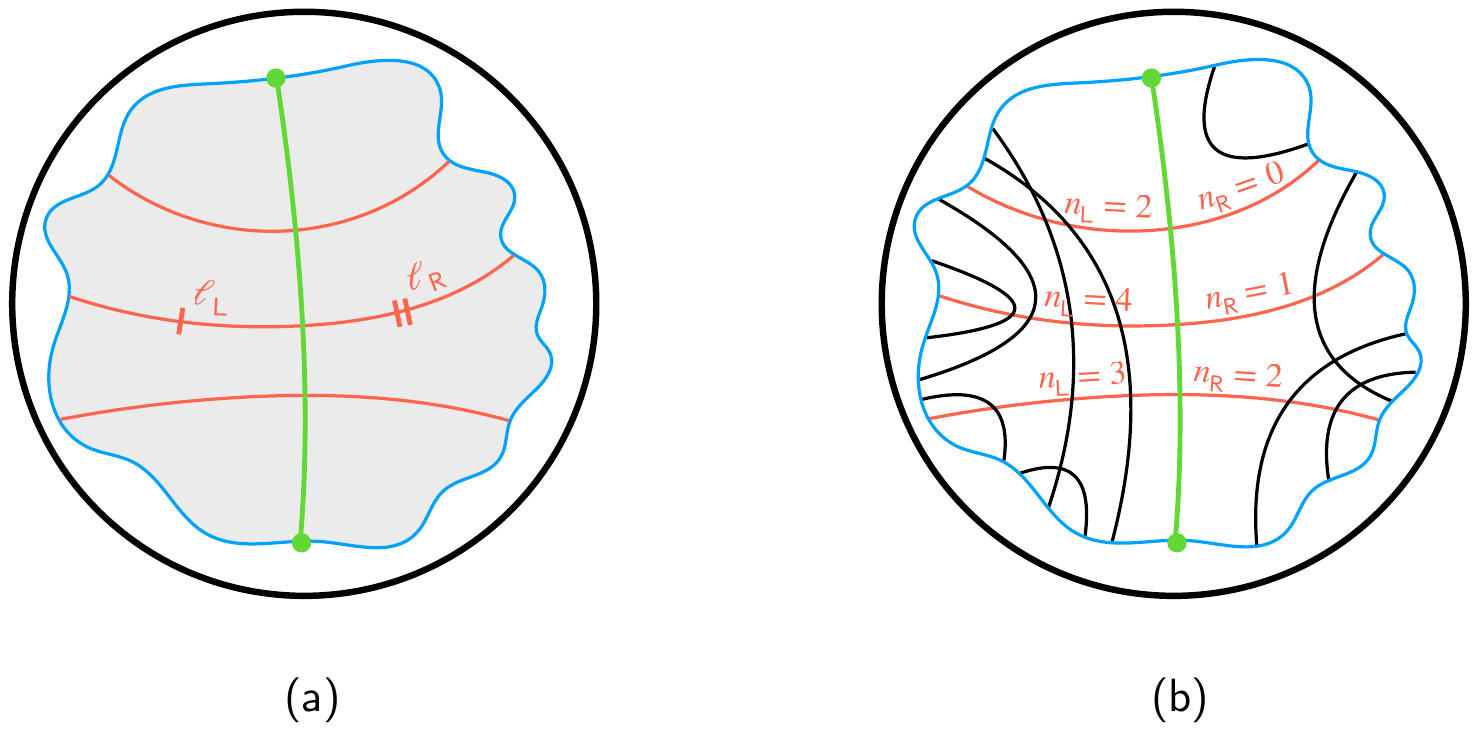}
    \end{center}
    \caption{(a) The Hilbert space of JT gravity with a particle (green) in the wormhole. The matter particle divides a geodesic from the left to right boundary into two pieces, with lengths $\ell_\lt$ and $\ell_\rt$. The disk partition function with two operator insertions is obtained by imposing $\ell_\lt= \ell_\rt = 0$ boundary conditions in the Euclidean past and future. (b) The Hilbert space of double scaled SYK with an operator insertion $M_s$ represented by a green chord. The Hilbert space is spanned by states labeled by two integers $n_\lt, n_\rt$ that are the left and right chord numbers.  }
    \label{hilbertParticle}
\end{figure}
Now we consider a state on the boundary that is obtained by acting on the TFD with an operator $M_s$ given by \nref{matter} at some point on the thermal circle. In the gravity description, we have inserted a particle somewhere in the wormhole. 
In the chord picture, we have inserted a new kind of chord, depicted in green in Figure \ref{hilbertParticle}.
With this additional chord type, cutting open a diagram along a red slice yields two integers $n_\lt$ and $n_\rt$, the number of chords to the left and right of the green chord. The appropriate generalization of \nref{tfd} is 
\begin{equation}
\begin{split}
\label{eq:1pstate}
    e^{-\tau_1 H} M_s e^{-\tau_2 H} \ket{\Omega} \;\; \cong \;\; e^{-\tau_1 H_\lt } e^{-\tau_2 H_\rt }\ket{n_\lt = 0, \, n_\rt = 0}_s.
\end{split}
\end{equation}
We have labeled the state \nref{eq:1pstate} by the subscript $s$ to indicate both the size and species of the matter particle. We drop this label when unambiguous.
Notice that we must now distinguish between the $H_\lt$ and $H_\rt$. Let's consider $H_\rt$: like before, it can create an open chord, but now there are two ways in which it can annihilate a chord:
\begin{figure}[H]
    \begin{center}
    \includegraphics[width = \columnwidth, trim={0, 3.6cm, 0, 6cm}, clip]{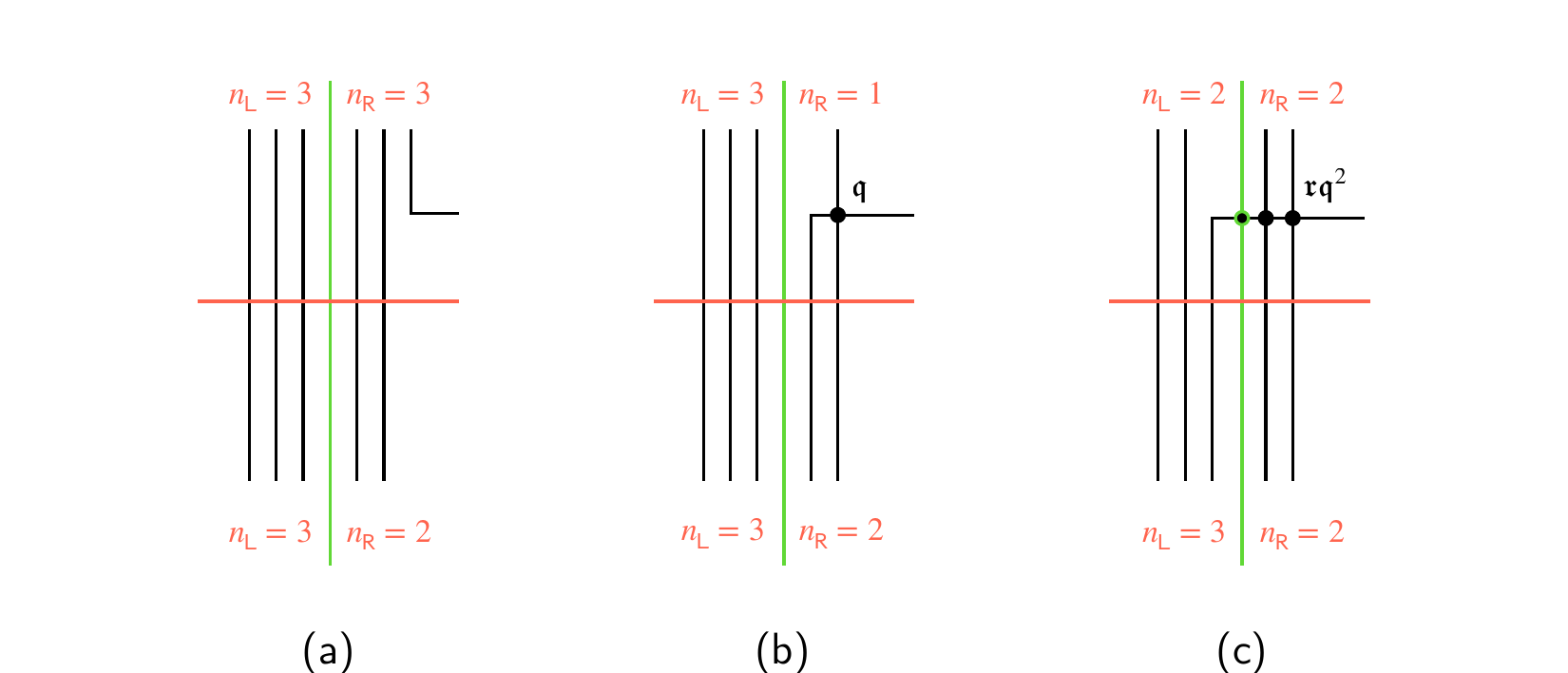}
    \end{center}
    \caption{Various processes can happen when we act with $H_\rt$. In (a) an extra chord is added. In (b) a chord from the right is removed. In this example, it crosses a chord. In (c) a chord from the left is removed. It crosses the matter particle and all $n_\rt$ chords. }
    \label{transferMatrix}
\end{figure}
\noindent The two ways of annihilating a chord give different $\alpha_\lt$ and $\alpha_\rt$ terms: %
\begin{equation}
\begin{split}
\label{tl1p}
    T_\rt \sqrt{\lambda} &=  \cre_\rt +  \alpha_\rt W_\rt + \alpha_\lt\rr \qq^{n_\rt}W_\lt,\\
      T_\lt  \sqrt{\lambda}  &=  \cre_\lt +  \alpha_\lt W_\lt + \alpha_\rt\rr \qq^{n_\lt}W_\rt  .
\end{split}
\end{equation}
There is a factor of $\rr \qq^{n_\rt}$ in the $T_\rt$ \nref{tl1p} since a left chord must cross all $n_\rt$ right chords before ending on the right. %
We have a similar expression for $T_\lt$.

As in the case with no free particles, we can define an inner product on the states $\ket{n_\lt, n_\rt}$. One follows the same reasoning of dividing the chord diagrams into a top, middle, and bottom region, and using the middle region to define an inner product.
For 1-particle states the new subtlety is that in the middle region, a black chord can cross the green chord. 
So states $\ket{n_\lt, n_\rt}$ are not orthogonal to states $\ket{n_\lt',n_\rt'}$ unless $n_\lt + n_\rt \ne n_\lt + n_\rt'$. One can derive a recursion relation, generalizing \nref{eq:smat}:
\begin{equation}
\begin{split}
\la{innerP1r}
\braket{n_\lt,n_\rt}{n_\lt', n_\rt'} &= 
\frac{1- \qq^{n_\lt}}{1-\qq} \braket{n_\lt-1,n_\rt}{n_\lt'-1, n'_\rt }%
+\rr \qq^{n_\lt} \frac{1- \qq^{n_\rt}}{1-\qq}\braket{n_\lt, n_\rt-1}{n_\lt'-1,n_\rt'}%
\\
\braket{n_\lt,n_\rt}{n_\lt', n_\rt'} &= 
\frac{1- \qq^{n_\rt}}{1-\qq} \braket{n_\lt,n_\rt-1}{n_\lt', n_\rt'-1}%
+\rr \qq^{n_\rt} \frac{1- \qq^{n_\lt}}{1-\qq}\braket{n_\lt-1, n_\rt}{n_\lt',n_\rt'-1}%
\end{split}
\end{equation}
In the first line, we again imagine deleting the left-most chord in the ket. The first term counts diagrams where the left-most chord stays to the left of the matter particle. The second term counts diagrams where the left-most chord crosses the matter particle (giving the factor of $\rr$) as well as all $n_\lt$ other left chords and then proceeds to cross some number of right chords. In the second line, we apply the same reasoning except we delete the right-most chord. The various processes are basically the same as in Figure \ref{transferMatrix}b and \ref{transferMatrix}c. 
As a sanity check, consider the $\rr \to 0$ limit. There is an extremely heavy particle which pinches off the thermal circle into two thermal circles. We see that in this regime, \nref{tl1p} and \nref{innerP1r} reduce to essentially two copies of the formulas in Section \ref{sec:pure}.

Supplementing these recursion relations are the ``boundary conditions'' when one of the chord numbers vanishes (compare with \nref{eq:smat_bd}):
\begin{equation}
\begin{split}
\la{eq:recursion_start}
\braket{0,n_\rt}{n_\lt', n_\rt'} &= \rr^{n_\lt'} \braket{n_\rt}{n_\lt' + n_\rt'}\\
\braket{n_\lt,n_\rt}{0, n_\rt'} &= \rr^{n_\lt} \braket{n_\lt + n_\rt }{ n_\rt'}
\end{split}
\end{equation}
On the RHS we have states with no matter particles, whereas on the LHS we have states with a single matter particle. This rule is explained by the following diagram:
\begin{figure}[H]
    \begin{center}
    \includegraphics[width = \columnwidth, trim={0, 7cm, 0, 7cm}, clip]{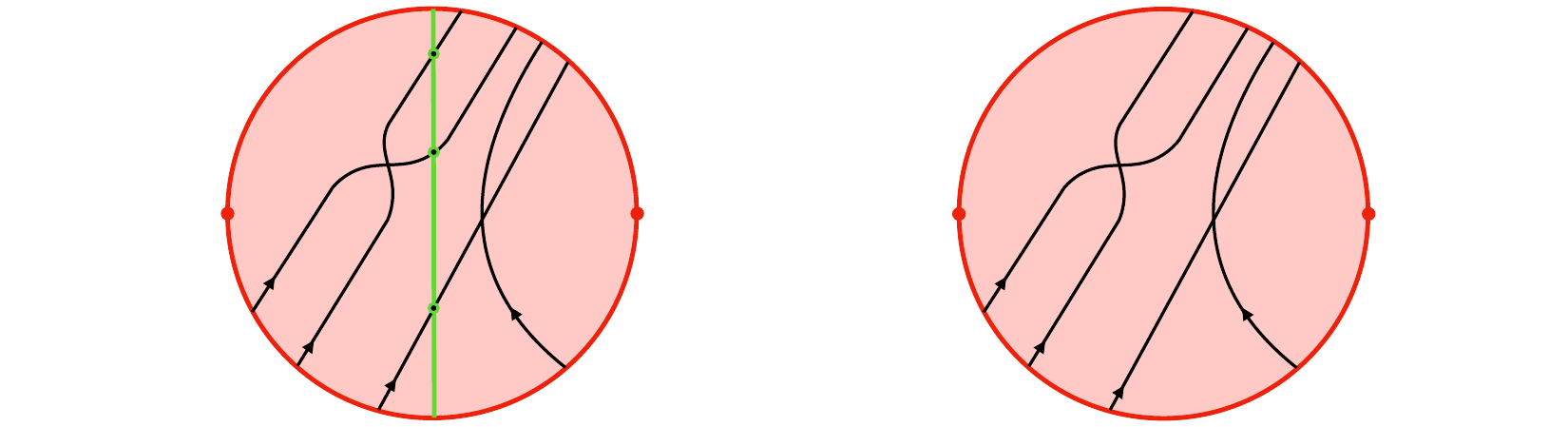}
    \end{center}
    \caption{The delete-a-matter-matter-chord rule \nref{eq:recursion_start}. These pink disks are the middle region in Figure \nref{inner}. }
    \label{fig:delete}
\end{figure}
\noindent On the LHS, we have a possible contribution to $\braket{3,1}{0,4}$. By deleting the matter chord, we get the contribution to $\braket{4}$ on the RHS. The deletion can be compensated for by a factor of $\rr^3$ coming from the $n_\lt'=3$ points where the green chord crosses the black chords. One can also obtain two more relations by permuting left and right in \nref{eq:recursion_start}.

\begin{figure}[t]
    \begin{center}
    \includegraphics[width=0.66\columnwidth]{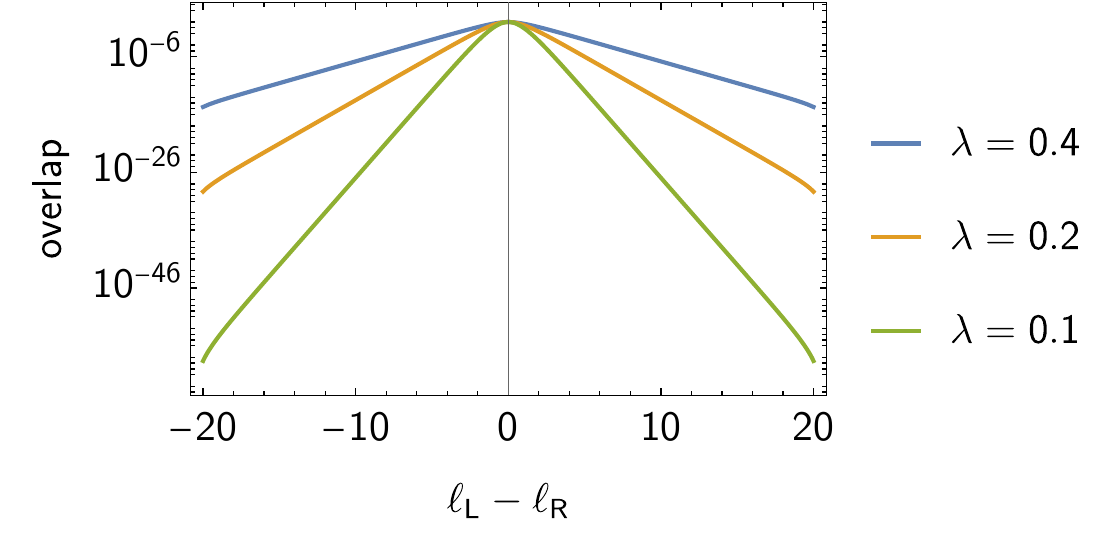}
    \vspace{-0.5cm}
    \end{center}
    \caption{The normalized overlap \nref{normalized_overlap_num} between the symmetric state $\ket{n, n}$ and the asymmetric state $\ket{n_\lt, n_\rt}$. We keep $\ell = \lambda n $ fixed and consider different $\lambda$. We set $\rr = 1/2$. If we were to plot the overlap as a function of $n$ instead of $\ell$, the curves would appear in reverse order.}
    \label{overlap}
\end{figure}

\begin{figure}[t]
    \begin{center}
     \hspace{0.9cm}\includegraphics[width=0.73\columnwidth]{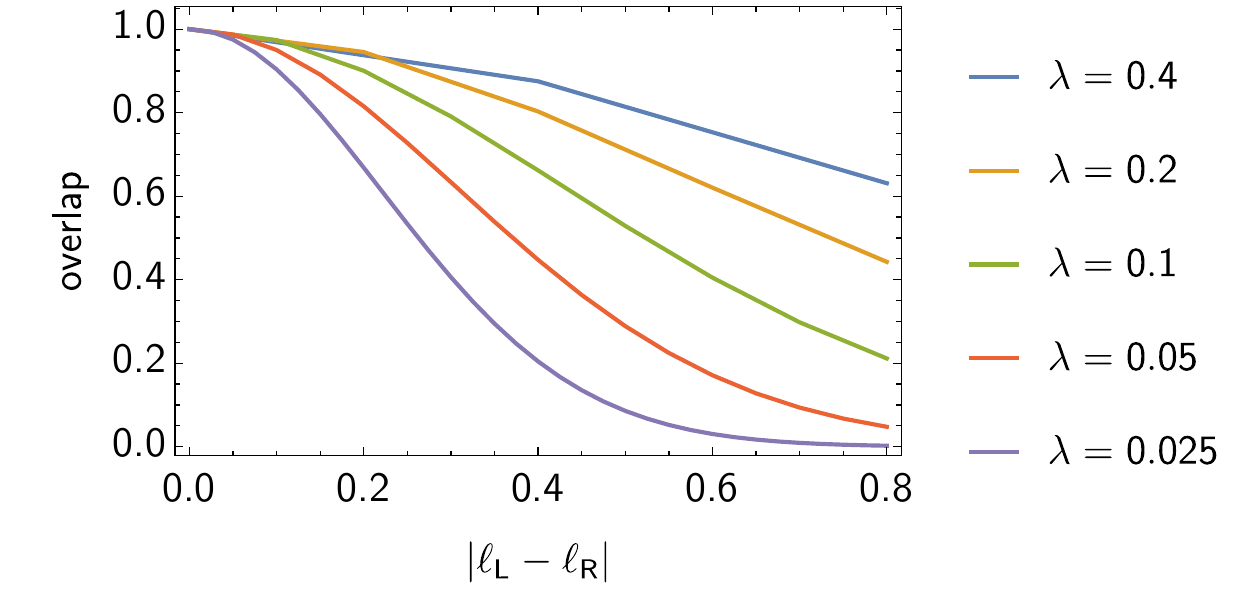}
    \vspace{-0.5cm}
    \end{center}
    \caption{Same overlap as in Figure \ref{overlap} but plotted on a smaller linear scale for clarity. The numerics suggest that if we hold $\ell_\lt - \ell_\rt$ fixed as $\lambda \to 0$, the overlap goes to zero. So in the JT regime, we expect that the states $\ket{\ell_\lt, \ell_\rt}, \ket{\ell_\lt',\ell_\rt'}$ are orthogonal. However, note that states that differ by only one unit of $n$ (e.g. a small change $\delta \ell \sim \lambda$) have a large overlap.  }
    \label{overlap2}
\end{figure}

This recursion relation is rather complicated because it involves two terms. However, it can be efficiently solved numerically using ``memoization.'' In Figure \nref{overlap}, we compute numerically the overlap of a symmetric state with asymmetric states (having the same total chord numbers), e.g.,
\begin{equation}
\begin{split}
\label{normalized_overlap_num}
  \mathsf{overlap} = \frac{\braket{n_\lt, n_\rt}{n, n}}{\sqrt{\braket{n,n} \braket{n_\lt, n_\rt }}}.
\end{split}
\end{equation}
An interesting feature is that in the triple scaling limit, we see numerical evidence that states with different $\ell_\lt$ or $\ell_\rt$ are orthogonal (assuming that the differences in lengths are held fixed as $\lambda \to 0$), see Figure \nref{overlap2}. This implies that we can treat $\ell_\lt, \ell_\rt$ as commuting operators in the JT limit.

It would be nice to solve the recursion relation exactly. We will not attempt this here, but note that a simple approximation to \nref{innerP1r} is to simply replace the second term with the first term: 
\begin{equation}
\begin{split}
\label{ratioApp}
\braket{n_\lt,n_\rt}{n_\lt', n_\rt'} \approx \frac{1-\qq^{n_\lt} + \rr \qq^{n_\lt} (1-\qq^{n_\rt}) }{1-\qq} \braket{n_\lt-1,n_\rt}{n_\lt'-1, {n'_\rt }}.
\end{split}
\end{equation}
This approximation works well in various regimes. First, if $n_\lt, n_\rt$ are large holding $\qq, \rr < 1$ fixed, we can neglect the second term entirely. This is the double scaled model at low energies. 
Second, we can consider the regime $\qq \to 1 $ but $\qq^n = e^{-\ell}$ and $\rr$ fixed\footnote{A previous version of this paper erroneously suggested that the inner product vanishes when $\Delta$ is held fixed.}. In this regime, $\braket{n_\lt-1,n_\rt}{n_\lt'-1, {n'_\rt }} \approx \braket{n_\lt,n_\rt-1}{n_\lt'-1, {n'_\rt }}$. This is the standard large $q$ SYK model at finite energies. In upcoming work \cite{linprep}, we will solve the inner product explicitly in this limit; at finite $\Delta$ the inner product is non-zero but as $\Delta \to \infty$ the inner product falls off quickly when $x= \lambda(n_\lt - n_\rt)/2$ is held finite.

\begin{figure}[t]
    \begin{center}
   \includegraphics[width = 0.8\columnwidth, trim={0, 0.2cm, 0, 0cm}, clip]
    {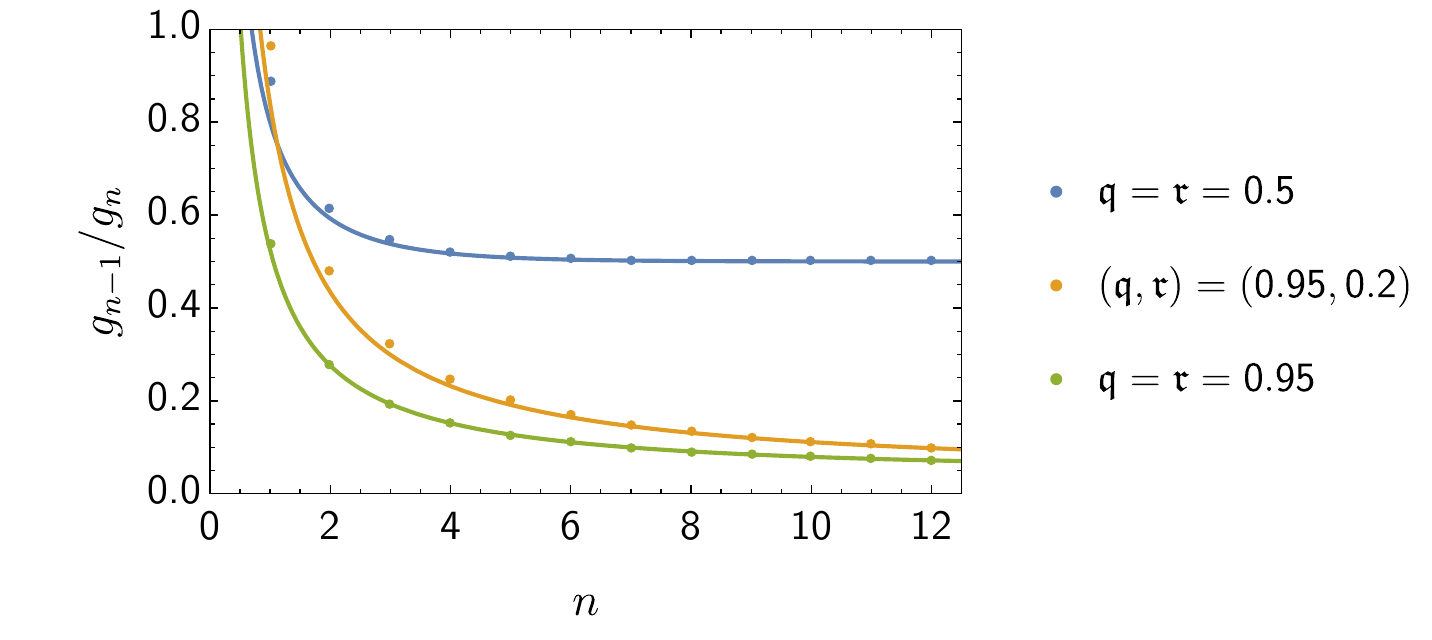}
    \vspace{-0.4cm}
    \end{center}
    \caption{Ratio of the inner product $g_{n-1}/g_{n} = \braket{n-1,n}/\braket{n,n}$.  The dots are computed by numerically solving the recursion relation \nref{innerP1r}. The solid lines are given by the approximation \nref{ratioApp}. As $\qq, \rr \to 1^-$, the approximation holds everywhere. For finite values $\qq, \rr < 1$, the approximation holds for large $n$.   }
    \label{overlap_ratio}
\end{figure}

Finally, this approximation holds in the triple-scaled limit (and $\Delta \to \infty$) where both effects are favorable. We compare this approximation to numerics in Figure \ref{overlap_ratio}. A word of caution: although the LHS and RHS of \nref{ratioApp} might be approximately equal for large values of $n$, replacing the entire recursion relation \nref{innerP1r} with \nref{ratioApp} does {\it not} give a good approximation to the true inner product, even at large $n$. Indeed, solving the recursion relation involves going to small $n$, where the approximation \nref{ratioApp} breaks down.

To make contact with JT gravity, we should ideally  perform a change of basis $T_\lt \to  -g^{1/2} T_\lt g^{-1/2}$. This would involve solving the recursion relation \nref{innerP1r}.
Instead, we conjugate $T_\lt \to -\tilde{g}^{1/2} T_\lt \tilde{g}^{-1/2}$ where $\tilde g$ satisfies \nref{ratioApp}. %
Then we take the $\lambda \to 0$ limit holding $\tl_\lt, \tl_\rt, \Delta$ fixed\footnote{A previous version of this paper had the transpose of this expression. I thank Edward Witten for pointing out this error.}:
\begin{equation}
\begin{split}
\label{hl1}
    \tl_\lt &=\ell_\lt + \log \lambda  , \quad     \tl_\rt =\ell_\rt +\log \lambda \\ %
    H_\lt  &\approx \frac{1}{ 2C } \lb  -\tilde{\pd}_\lt^2 +\Delta  e^{-\tl_\lt } + \lp \tilde{\pd}_\lt -  \tilde{\pd}_\rt\rp  e^{-\tl_\lt } + e^{-\tl_\lt -\tl_\rt}\rb \\
    H_\rt  &\approx \frac{1}{ 2C } \lb  -\tilde{\pd}_\rt^2 +\Delta  e^{-\tl_\rt } +  \lp \tilde{\pd}_\rt -  \tilde{\pd}_\lt\rp  e^{-\tl_\rt }+ e^{-\tl_\lt -\tl_\rt}\rb. \\
\end{split}
\end{equation}
We have also shifted both Hamiltonians by an amount $E_0$ in \nref{eq:tripSHam} so that the ground state has zero energy. 
In JT gravity, we expect that $\tl_\lt$ has the interpretation of the (renormalized) length of the segment of $\gamma_{\lt \rt}$ which extends from the left boundary to the matter particle, where $\gamma_{\lt \rt}$ is the geodesic that extends from the left to right boundaries\footnote{As a sanity check, note that if we restrict to symmetric configurations $\ell_\lt(\tau) = \ell_\rt(\tau) = \ell/2$,
$ H_\lt + H_\rt= \frac{1}{C} \lp  k^2 +   e^{-\ell} + \Delta e^{-\ell/2}\rp  .$
The effective potential $V(\ell)= 2e^{-\ell} +\Delta e^{-\ell/2} $ was derived in \cite{Maldacena:2018lmt} using classical considerations. }, see Figure \nref{hilbertParticle}. In the remainder of this subsection, we will point out some properties of \nref{hl1} from the viewpoint of JT gravity.

The expressions \nref{hl1} are the generalization of the Liouville Hamiltonian \nref{jt2} to the case where there is a single particle in the bulk. Note that the Hamiltonians act on a 4 dimensional phase space. In pure JT gravity (no matter), one can think of the two dimensional phase space as arising from the complex length $\beta + i T$ of the thermal circle ($\beta$ controls the energy whereas $T$ is the Lorentzian time). With a single matter particle, the classical solutions are 2 thermal circles glued together \cite{Kourkoulou:2017zaj, Goel:2018ubv}. 
	Thus we expect a 2 complex lengths, or a 4 dimensional phase space, which agrees with the above analysis.

A peculiarity is that the Hamiltonians defined in \nref{hl1} are not Hermitian with respect to the naive inner product $\braket{\psi}{\chi} = \int d\tl_\lt d\tl_\rt \, \psi^* \chi$. Since we performed a change of basis using $\tilde{g}$ defined by \nref{ratioApp} instead of $g$ defined by \nref{innerP1r}, there was no guarantee that the inner product would be the canonical one. 
However, for large $\tl_\lt, \tl_\rt$ we do expect that the inner product should agree with the canonical one\footnote{We attempted to study the inner product numerically in the triple scaling regime. Solving \nref{ratioApp} instead of \nref{innerP1r} seems to give a good approximation up to an overall factor that is independent of $\tl_\lt, \tl_\rt$ when these renormalized lengths are large.}.
Indeed, note that $H_\lt$ has the form of a scattering Hamiltonian in the $\tl_\lt$ variable, and therefore its eigenvalues are determined by the asymptotic region $\tl_\lt \to \infty$ where we have an incoming/outgoing plane wave. In this region, $H_\lt \propto k_\lt^2$ which is manifestly real.\footnote{Another consequence of the $i$ is the following. Note that $i e^{-\tl_\lt} k_\lt = i k_\lt  e^{-\tl_\lt} + e^{-\tl_\lt}$. Hence we could choose the opposite ordering if we shift $\Delta \to \Delta - 1$ by a real amount. } We thus expect that $H_\lt, H_\rt$ are Hermitian with respect to the inner product $g$. For our purposes below, we will not need this inner product, although it would be nice to work it out.

Notice also that since $H_\lt$ and $H_\rt$ are separately conserved, the ingoing momentum of the particle is opposite the outgoing momentum $(k_\lt, k_\rt) \to -(k_\lt, k_\rt).$ We expect the system to be integrable since there are two conserved charges and two coordinates. It would be interesting to solve this quantum mechanical scattering problem, which should give the OTOC/6j symbol \cite{Mertens:2017mtv}. More explicitly, we can define a perturbed TFD by imposing boundary conditions $\ell_\lt \to -\infty, \ell_\rt \to -\infty$ in the Euclidean past and future, see \ref{hilbertParticle}. The OTOC is then
\begin{equation}
\begin{split}
\label{}
 \tr[O_\Delta  e^{-\tau_1  H} O_{\Delta'} e^{-\tau_2 H} O_\Delta e^{-\tau_3 H} O_{\Delta'} e^{-\tau_4 H} ]\propto \lim_{\tl_c \to -\infty} \bra{ \tl_c,  \tl_c}  e^{- \tau_1 H_\lt -\tau_4 H_\rt} e^{-\Delta' \tl }e^{- \tau_2 H_\lt -\tau_3 H_\rt} \ket{\tl_c,  \tl_c}.
\end{split}
\end{equation}
Here $\ket{\ell_c, \ell_c}$ is a delta-function in position space $\psi(\ell_\lt,\ell_\rt) = \delta(\ell_\lt-\ell_c) \delta(\ell_\rt-\ell_c)$. The dependence on $\Delta$ enters via $H_\lt, H_\rt$. %

An interesting property of JT gravity with arbitrary matter is that there exists a gravitational algebra \cite{Harlow:2021dfp} which contains the left/right Hamiltonians and the total length operator. 
In our conventions\footnote{In the conventions of \cite{Harlow:2021dfp} $ 2\phi_b = C$ and $\tilde{L} = \tl + \log(4).$},
\begin{equation}
\begin{aligned}
\la{harlow_wu}
[{\tl}, k_{\lt/\rt }] &=i \\
[k_{\lt}, k_{\rt}] &=0\\
[H_{\lt}, H_{\rt} ] &=0 \\
-i[{\tl}, H_{\lt/\rt} ]  &=\frac{k_{\lt/\rt}}{C} \\
-i[{k}_{\lt/\rt}, H_{\lt/\rt}]  &=H_{\lt/\rt}-\frac{{k}_{\lt/\rt}^{2}}{2C } \\
-i[{k}_{\lt/\rt}, H_{\rt/\lt}] &= \frac{e^{-\tl}}{2C} \\
\end{aligned}
\end{equation}
In \cite{Harlow:2021dfp} the algebra was derived in the classical approximation using Poisson brackets. In appendix \ref{app:quantumwu}, we give a quantum derivation of the above formulas.

Specializing to the case of a single particle in the bulk, we obtain a concrete representation of this algebra in terms of a Hilbert space of wavefunctions with two coordinates $\psi(\ell_\lt, \ell_\rt)$. Indeed, one can check \nref{harlow_wu} using the expressions \nref{hl1} together with
\begin{equation}
\begin{split}
\label{gravalg1}
\tl &=\tl_L + \tl_R  \\
k_\lt &= -i\pd/\pd{\ell_\lt} \\
k_\rt &= -i\pd/\pd{\ell_\rt}.
\end{split}
\end{equation}
The most non-trivial commutation relation to check is actually $[H_\lt, H_\rt] = 0$. The last relation in \nref{harlow_wu} determines the $\ell_\rt$ dependence in $H_\lt$.
Indeed, with a little guesswork one can obtain the form of \nref{hl1} just using the commutation relations.

An important fact about the gravitational algebra is that it contains an \slt\, subalgebra, which are the symmetry generators that move matter around in the wormhole \cite{Lin:2019qwu, Harlow:2021dfp}:
\begin{equation}
\begin{split}
\label{slt}
  L_0 &= i(k_\lt - k_\rt), \\
  L_+ &= 4 C e^{\tl/2} (H_\lt - k_\lt^2 -e^{-\tl} )\\
  L_- &= 4 C e^{\tl/2} (H_\rt - k_\rt^2 -e^{-\tl} ).\\
\end{split}
\end{equation}
\noindent
Let $x$ be the relative position of the matter particle from the mid point. Then we can write a simple expression for the generators:
\begin{equation}
\begin{split}
\label{slt2}
x &= (\tl_\lt - \tl_\rt)/2\\
  L_0 &= -i\pd_x, \quad   L_\pm =   \lp  \Delta \pm   \pd_x \rp e^{\mp x}  \\
\end{split}
\end{equation}

\noindent
These satisfy the commutation relations $[L_m,L_n] = i(m-n)L_{m+n}$. The operator $\tilde P = L_0$ is the momentum generator, the global energy $\tilde{E} = (L_+ + L_-)/2 $ and the boost $\tilde{B} = (L_+ -L_-)/2$, where we have converted to the notation of \cite{Lin:2019qwu}. For each value of $\Delta$, \nref{slt} defines a concrete representation of the \slt\:algebra which acts on wavefunctions of two variables.
The expression for $L_0$ is particularly simple. In the double scaled model, it can be realized by an operator which shifts the position of a particle by creating a chord to the right and annihilating a chord to the left. %
We can also define the Casimir
\begin{equation}
\begin{split}
\label{}
  \mathcal{C} =    - L_0^2 + \hf \lp L_+ L_- +L_- L_+\rp  =  \Delta (\Delta-1).
\end{split}
\end{equation}
Given boundary expressions for the total chord number, one can use the above expressions to construct the symmetry algebra on the boundary. We will return to this point in Section \ref{sec:dsalg}.
Since $\pd_x$ annihilates $\tl$, it is trivial to check that the length $\tl$ commutes with $L_n$ in \nref{slt2}. This is true for general matter and follows directly from just the algebra \nref{slt2}, see \cite{Harlow:2021dfp}. Thus the representations of \slt\: decomposes into infinitely many discrete series parameterized by $\tl$. Note that the continuous series is not present in this discussion, since these are the matter generators and the continuous series is associated with the Schwarzian particle.

\subsection{Generalization to multiple particles \la{multiparticles}}
 \begin{figure}[t]
    \begin{center}
	\vspace{-3.5cm}
    \includegraphics[width=\columnwidth]{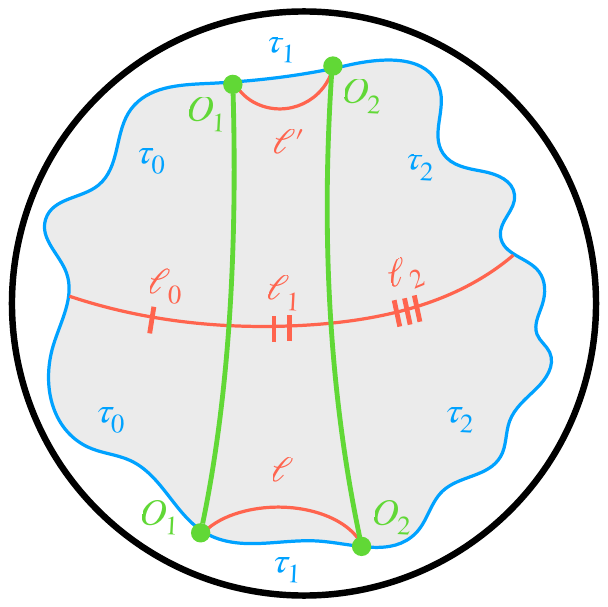}
    \vspace{-4cm}
    \end{center}
    \caption{JT gravity with 2 particles in the wormhole. Cutting open the diagram on a geodesic gives a bulk Hilbert space with 3 lengths $\ell_0, \ell_1, \ell_2$. To evaluate this diagram, one can cut the diagram along the asymptotic geodesics (lengths $\ell_1 = \ell$ and $\ell_1 = \ell'$) in the Euclidean past and future. One then sews on a cap $\braket{\ell}{\tau_1}$. }
    \label{hilbertParticle2}
\end{figure}

Now we consider generalizing the above discussion to $m$ particles in the bulk. 
To specify a state in the bulk Hilbert space we must specify $m+1$ chord numbers,
\begin{equation}
\begin{split}
\label{}
  \ket{n_0, n_1, \cdots, n_m}_{s_1, s_2, \cdots, s_m}
\end{split}
\end{equation}
Here we have also labeled the state by the sizes of the matter operators $s_1, s_2, \cdots, s_m$.
Equivalently, we can parameterize these states by fixing the total number chords $n_\text{tot}$ (of any type). We can think of $n_\text{tot}$ as the length of a lattice with $n_\text{tot}$ sites.
On each site, one can either have no matter excitation (a black chord) or a particle of some type with size $s_i$. The Hilbert  of each lattice site has a dimension that is determined by the number of chord species.\footnote{In the double scaled limit, we can consider infinitely many species of random operators $M_s$ \nref{matterop}, so the dimension of each lattice site is infinite. Of course, we can consider a subspace of the bulk Hilbert space where we act only with a finite number of species of $M_s$, in which case the dimension would be finite.} So in summary, we can think of this theory as a lattice field theory, where the total number of lattice sites (the length of the lattice) is dynamical:
\begin{equation}
\begin{split}
\label{}
  \mathcal{H}_\mathsf{bulk} = \bigoplus_{n_\mathsf{tot} \in \mathbb{Z}^+ } \mathcal{H}_\mathsf{lattice}(n_\text{tot})
\end{split}
\end{equation}
This resembles the Hilbert space of a bulk quantum field theory except that there is an explicit UV regulator set by $\lambda$.

Next we consider expressions for $H_\lt$ and $H_\rt$ when we have $m$ particles in the bulk. We will write expressions that hold when the particles are of the same or different species, with sizes $s_1, \cdots, s_m$ (equivalently with interaction vertices $\rr_1, \cdots, \rr_m$).
The left and right transfer matrices are
\begin{equation}
\boxed{
\begin{split}
\label{generalT}
T_\lt \sqrt{\lambda} &= e^{i \lambda k_0} + \sum_{i=0}^m e^{-i \lambda k_i}\lp \frac{1-e^{-\ell_i}}{1-\qq}\rp \prod_{j<i} \rr_j e^{-\ell_j} ,\\
  T_\rt \sqrt{\lambda} &= e^{i \lambda k_m} + \sum_{i=0}^m e^{-i \lambda k_i}  \lp \frac{1-e^{-\ell_i}}{1-\qq}\rp  \prod_{j>i} \rr_j e^{-\ell_j} .\\
\end{split}
}
\end{equation}
Here $k_i$ are defined to be the conjugate momenta to $\ell_i$.  For $T_\lt$, we define the product over $j$ to be 1 when $i=0$, and similarly for $T_\rt$ the product is 1 when $i=m$.
For general states with $m$ particles in the interior (potentially of different sizes), we have
\begin{equation}
\begin{split}
\label{}
	\braket{n_0, \cdots, n_m}{n_0', \cdots, n_m'} &=  \frac{1- \qq^{n_0}}{1-\qq} \braket{n_0-1, \cdots, n_m}{n_0'-1, \cdots, n_m'}  \\
	&+\sum_{k=1}^m \qq^{\sum_{i<k} n_i} \rr_1 \cdots \rr_k \frac{1- \qq^{n_k}}{1-\qq} \braket{n_0, \cdots, n_k-1, \cdots,  n_m}{n_0'-1, \cdots  n_m'}
\end{split}
\end{equation}
This recursion relation involved deleting the left-most chord.  We could also obtain a similar recursion relation involving the right most chord. 
By iterating this recursion relation, we can get down to states of the form $\braket{0,n_1,\cdots, n_m}{n_0',n_1' \cdots, n_{m}' }$. Then we can generalize the delete-a-matter-chord rule (see \nref{eq:recursion_start} and Figure \ref{fig:delete}):
\begin{equation}
\begin{split}
\label{}
  \braket{0,n_1,\cdots, n_m}{n_0',n_1' \cdots, n_{m}' } = \rr_1^{n_0'} \braket{n_1, \cdots, n_m}{n_0'+n_1',n_2' \cdots, n_m'}
\end{split}
\end{equation}
So we have reduced the problem to studying norms of states with $m-1$ particles in the interior. In this way, we can recursively build norms.

Then performing the change of basis and taking the triple scaling limit, %
\begin{equation}
\boxed{
\begin{split}
\label{eq:generalSch}
 H_\lt &= \frac{1}{2C} \lb  k_0^2 + e^{-\tl }  - i\sum_{i=1}^m   (k_i - k_{i-1}) e^{-\sum_{j<i} \tl_j}+   \sum_{i=1}^m \Delta_i e^{-\sum_{j<i} \tl_j} \rb \\
  H_\rt &= \frac{1}{2C} \lb  k_m^2 + e^{-\tl }  + i\sum_{i=1}^m   (k_i - k_{i-1}) e^{-\sum_{j\ge i} \tl_j}+   \sum_{i=1}^m \Delta_i e^{-\sum_{j \ge i} \tl_j} \rb  
\end{split}
}
\end{equation}
We have defined renormalized lengths $\tl_0 = \ell_0 + \log \lambda $ and $\tl_m + \log \lambda $. The other lengths $\tl_1 = \ell_1, \cdots , \tl_{m-1} = \ell_{m-1}$ are {\it not} renormalized. In JT gravity, this makes sense: the lengths $\ell_1, \cdots \ell_{m-1}$ are geodesic segments that do not approach the asymptotic boundary. For $m>1$, this means $\tl_0$ and $\tl_m$ play a distinguished role in the scattering problem. They are the only lengths that have a quadratic kinetic term. Also notice that all non-Hermitian terms are suppressed by $e^{-\tl_0}$ in $H_\lt$ and $e^{-\tl_m}$ in $H_\rt$.

One can again check the gravitational algebra \nref{harlow_wu} using
\begin{equation}
\begin{split}
\label{}
\tl  & =\tl_0 + \ell_1+\cdots+ \ell_{m-1} + \tl_m \\
k_\lt &= k_0 =  -i\pd/\pd {\tl_0} \\
k_\rt &= k_m= -i\pd/\pd {\tl_m} \\
\end{split}
\end{equation}
Notice once again that in $H_\lt$ the only dependence on the right length $\tl_m$ is via $\tl_\tot$. This is required if the last relation \nref{harlow_wu} is to be satisfied. In addition, note that $(k_i - k_{i-1})$ commutes with $\tl$, which is required for the fourth relation in \nref{harlow_wu}.

A new complication for $m>1$ is that the wormhole state is not obtained by simply taking all the chord numbers to zero.
Instead, one takes $n_0, n_m = 0$ but sums over the remaining intermediate $n_i$, ``gluing'' together pieces of the empty wormhole, see Figure \ref{hilbertParticle2}. For $m=2$, this means that the analog of \nref{tfd} is 
\begin{equation}
\begin{split}
\label{}
 &e^{-\tau_0 H} M_s  e^{-\tau_1 H}  M_s e^{-\tau_2 H} \ket{\Omega}  \cong \ket{\tau_0, \tau_1,\tau_2} \\
&\ket{\tau_0, \tau_1,\tau_2} = \sum_\ell \int  dE_1   e^{-\tau_0 H_\lt - \tau_1 E_1 - \tau_2 H_\rt} \ket{\ell_0 =0, \ell_1 = \ell,\ell_2 =0}\braket{\ell}{E},\\
\end{split}
\end{equation}
Here $\braket{\ell}{E} $ are the energy eigenstates of the empty wormhole Hamiltonian \nref{eq:tripSHam}, which are explicitly given in equation 2.16 of \cite{Berkooz:2018jqr} in terms of $\qq$-Hermite polynomials. This is the appropriate generalization of the Bessel K functions in JT gravity. %
In the context of JT gravity, such cut-and-paste constructions were discussed in \cite{Saad:2019pqd}, see also \cite{Penington:2019kki}. Here, we are showing that the same kind of cut-and-paste constructions also works in the double scaled theory using chord diagrams. 

Like in the $m \le 1$ cases, we can also compute boundary correlation functions in this formalism. For example,
\begin{equation}
\begin{split}
\label{}
\tr( O_\Delta e^{-\tau_0 H } O_1 e^{-\tau_1 H} O_2 e^{-\tau_2 H} O_\Delta e^{-\tau_2 H} O_2 e^{-\tau_1 H} O_1   ) =   \langle{\tau_0,\tau_1,\tau_2}| e^{-\Delta \tl_\tot}|\tau_0,\tau_1,\tau_2\rangle 
\end{split}
\end{equation}
Such correlation functions probe the length of the wormhole with $m$ particles in the interior\footnote{It would be interesting to find the $\nn = 2$ generalization of \nref{eq:generalSch}. For $m=0$, $H_\lt = H_\rt$ and we get $\nn=4$ Liouville quantum mechanics \cite{LongPaper,Lin:2022rzw} with a single bound state ($\hat{q}= 1$). For $m>1$, we expect that $H_\lt$ and $H_\rt$ would be $\nn = 2$ supersymmetric Hamiltonians, so that we still get 4 anti-commuting supercharges in total. For $m=1$ we expect a single bound state satisfying $H_\lt = H_\rt = 0$. For $m \ge 2$, it seems plausible that there are infinitely many bound states, corresponding to the different primaries one can make by combining two matter particles $O_1$ and $O_2$. The $m=2$ generalization would allow us to study the intermediate length $\ell_1$ between the two matter particles in the interior.}.

 \begin{figure}[t]
    \begin{center}
	\vspace{-5cm}
    \includegraphics[width=\columnwidth]{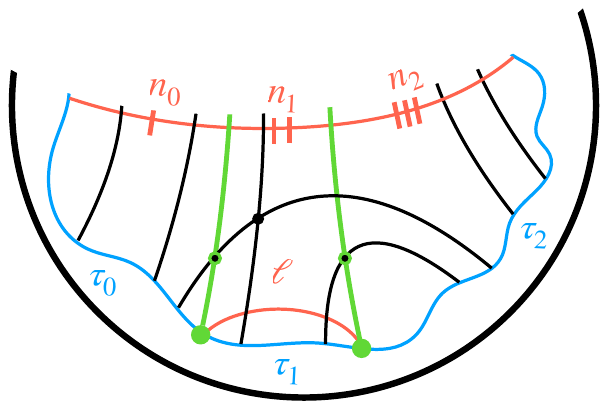}
    \vspace{-4cm}
    \end{center}
    \caption{The bulk Hilbert space with 2 particles in the interior in the chord picture. We can cut the chord diagram along in the Euclidean past $\ell$. This is an ``inital state'' in the Euclidean past with $n_0 = 0, n_1 = \ell/\lambda,$ and $n_2 = 0$. The $n_1$ wavefunction can be obtained by summing chord diagrams with no particles, as in Section \ref{sec:pure}. } %
    \label{hilbertParticleChord}
\end{figure}

\section{The bulk-to-boundary map \la{bulkBd}}
\subsection{A recursive reconstruction of chord states}
Besides describing the bulk Hilbert space, we would like to know the bulk-to-boundary holographic map.
Said differently, what boundary states correspond to the bulk states $\ket{n}, \ket{n_\lt, n_\rt}_s $, etc.? Let's start with the wormhole states $\ket{n}$ with no matter. A guess is simply $H^n \ket{\Omega}$ where $\ket{\Omega}$ is the maximally entangled state (the infinite temperature TFD). We are on the right track, but note that the state $H^n \ket{\Omega}$ has non-zero overlaps $\bra{m} H^n \ket{\Omega}$ for $m < n$ (and with the same parity as $n$). To obtain the states with definite chord number, one should perform the Gram-Schmidt process\footnote{I thank Douglas Stanford for suggesting this and pointing out the connection to Krylov complexity.} on the vectors $\{\ket{\Omega}, H \ket{\Omega}, H^2 \ket{\Omega}, \cdots  \}$ which gives $\{ \ket{0}, \ket{1}, \cdots \ket{n} \}$. One can explicitly work this out for the first few chord states: 
\def\th{\tilde{H}}
\begin{equation}
\begin{split}
\label{orthogonal}
  \ket{0} = \ket{\Omega}, \quad  \ket{1} &= \th \ket{\Omega}, \quad \ket{2} = \th^2 \ket{\Omega} - \frac{\ev{\th^2}{\Omega}}{\braket{0}} \ket{0}, \quad \ket{3} = \th^3 \ket{\Omega} - \frac{ \ev{\th^4}{\Omega}}{\braket{1} }  \ket{1}, \\
  \th &= \sqrt{\lambda} H\\
\end{split}
\end{equation}
The $=$ sign in the first line of \nref{orthogonal} really means ``$\cong$ under the bulk-to-boundary map'' but we leave this distinction implicit from now on. In the above, we can take all operators to be acting on the left side of the 2-sided state $\ket{\Omega}$.
Using the transfer matrix given in \nref{transferBerkooz}, one can compute the overlaps by diagonalizing $T$ \cite{Berkooz:2018jqr}. The energy eigenstates are scattering states with momentum $k = \theta/\lambda$ and energies given by \nref{eigenH}:
\begin{equation}
\begin{split}
\la{powers_of_H}
  \bra{\Omega} \th^n \ket{\Omega} &= \bra{0} T^n \ket{0} = \int_0^\pi \frac{d \theta }{2 \pi }  (\qq;\qq)_\infty (e^{\pm 2 i \theta}; \qq)_\infty   E(\theta)^n, \quad E(\theta) =  \frac{2 \cos \theta }{\sqrt{1-\qq }} \\
\end{split}
\end{equation}
The $\pm$ symbol in the first line means a product over the different signs, e.g., $(e^{\pm 2 i \theta}; \qq) = (e^{ 2 i \theta}; \qq)(e^{- 2 i \theta}; \qq)$. %
Despite the complicated appearance, this integral produces a polynomial in $\qq$ of order with integer coefficients\footnote{One can check this in Mathematica by Taylor expanding the integrand in $\qq$ to large order and then performing the integral \nref{powers_of_H}.}, as expected from the explicit expression for $T$. For small values of $n$, one can simply enumerate the chord diagrams instead of performing this integral. This gives
 \begin{equation}
\begin{split}
\label{}
  \ket{0} = \ket{\Omega}, \quad  \ket{1} = \th \ket{\Omega}, \quad \ket{2} = \th^2 \ket{\Omega} -\ket{\Omega}, \quad \ket{3} = \th^3 \ket{\Omega} -  (2 + \qq ) \th   \ket{\Omega}
\end{split}
\end{equation}
Using \nref{powers_of_H}, one can also check that the norm of the states (using the boundary expressions) are given by \nref{eq:smat_sol}. Thus the bulk-to-boundary map preserves the inner product.
	Note that this procedure is similar to the procedure used to define Krylov complexity \cite{Parker:2018yvk, Rabinovici:2020ryf, Balasubramanian:2022tpr}. 
One can say that for the case with no matter in the wormhole, the Krylov complexity agrees with the chord number (or length).

An equivalent way of explaining the above procedure is to examine the form of $T$ in \nref{transferBerkooz}. $T$ can both increase or decrease the chord number. So starting with a state $\ket{0}$ and acting on it with $T,\,  T^2, \,T^3,\, \cdots$, we generate states with increasing chord number. Since we know the explicit form of $T$, we can compute explicitly $T^n \ket{0}$ and subtract off appropriate amounts of the states $\ket{n-2}, \ket{n-4}$, etc.

Now let us work out some expressions for the boundary states that are dual to 1-particle states $\ket{n_\lt, n_\rt}$. For the 1-particle states, we can organize them in a rectangular array $\ket{n_\lt ,n_\rt }$:
\begin{equation}
\begin{split}
\label{array_states}
  \begin{bmatrix}
\ket{0,0} \ra & \ket{0,1} \ra  & \ket{0,2} \ra &  \cdots  \\
\hspace{-0.45cm}\downarrow & \hspace{-0.45cm}\downarrow & \hspace{-0.45cm}\downarrow  \\ 
\ket{1,0} \ra & \ket{1,1} \ra & \ket{1,2} \ra & \cdots\\
\hspace{-0.45cm}\downarrow & \hspace{-0.45cm}\downarrow & \hspace{-0.45cm}\downarrow  \\ 
\ket{2,0} \ra & \ket{2,1} \ra & \ket{2,2} \ra & \cdots \\
\hspace{-0.45cm}\downarrow & \hspace{-0.45cm}\downarrow & \hspace{-0.45cm}\downarrow  \\ 
\hspace{-0.45cm}\vdots & \hspace{-0.45cm}\vdots & \hspace{-0.45cm}\vdots & \hspace{-0.45cm}\ddots & \\
\end{bmatrix}
\end{split}
\end{equation}
One can define the $\ket{n_\lt, n_\rt}$ inductively by imagining that we already know all the states above it or to its left. Then by acting with $H_\lt$ we can generate a new state $\ket{n_\lt+1, n_\rt }$, see equation \nref{tl1p}. $H_\lt$ can also decrease $n_\lt$ or $n_\rt$ by 1, but those states are already known, so we can subtract them off. (Importantly, the expression for $T_\lt$ in \nref{tl1p} does not contain an $\alpha_\rt^\dagger$.) Similarly acting with $H_\rt$ we can generate $\ket{n_\lt,n_\rt+1}$. %
Explicitly,%
\begin{equation}
\begin{split}
\label{ex2}
  \ket{0,0} &= M_s \ket{\Omega}, \quad \ket{1,0} = \th  M_s  \ket{\Omega}, \quad \ket{0,1} = M_s \th   \ket{\Omega},\\
  \ket{1,1} &= \th M_s \th \ket{\Omega} - \rr \ket{0,0}, \quad \ket{2,0} = \th^2 M_s  \ket{\Omega} - \ket{0,0}, \quad \ket{0,2} =  M_s \th^2  \ket{\Omega} - \ket{0,0},\\
  \ket{2,1} &= \th^2 M_s \th  \ket{\Omega} - \rr (2+ \qq) \ket{1,0} - (1 + \rr^2 + \rr^2 \qq) \ket{0,1}\\
  \ket{3,0} &= \th^3 M_s  \ket{\Omega} - (2 + \qq) \ket{1,0}
\end{split}
\end{equation}
We can also view the above procedure as a Gram-Schmidt-like process. However, since the states $\ket{n_\lt, n_\rt}$ are not orthogonal to states $\ket{n_\lt', n_\rt'}$, it is a modified Gram-Schmidt procedure. In particular, we only enforce that the states $\ket{n_\lt, n_\rt}$ are orthogonal to states $\ket{n_\lt', n_\rt'}$ where $n_\lt ' + n_\rt' < n_\lt + n_\rt$. In practice, this makes life a bit easier, because we do not need to worry about how to order states with the same total chord number. For example, the state $\ket{2,1}$ in this approach should be orthogonalized with respect to $\ket{0,1}$ and $\ket{1,0}$ but not $\ket{2,1}$, see \nref{ex2}.

A convenient way of organizing the computation is to imagine that we have already derived the boundary expressions for the wormhole states with no matter $\ket{n}$. Then by acting with $(M_s)_\rt$, we generate $\ket{n,0}$. This gives us all the states in the first column of \nref{array_states}. Then by acting with $H_\rt$, we generate the next column. Every time we generate a new column, we orthogonalize new states with respect to states all states above and to the left of it. Once we are a finished generating a column, again act with $H_\rt$. %

To obtain a general formula for the orthogonalized states, we need to compute overlaps $\braket{n_\lt, n_\rt}{n_\lt',n_\rt'}$. These overlaps are related to 2-point functions on the disk, which were in turn already obtained by \cite{Berkooz:2018jqr}:
\begin{equation}
\begin{split}
\label{}
\bra{\Omega}  \th^{n_1} M_s \th^{n_2} M_s \ket{\Omega} %
  & = \int_{0}^{\pi} \prod_{j=1}^{2}\left[\frac{d \theta_{j}}{2 \pi}  (\qq;\qq) \lp e^{\pm 2 i \theta_{j}}; \qq\rp   E(\theta_j)^{n_{j}}\right] \frac{\left(\rr^{2} ; \qq\right) }{\left(\rr e^{i\left(\pm \theta_{1} \pm \theta_{2}\right)} ; \qq\right)}
\end{split}
\end{equation}
Once again, this complicated expression evaluates to a polynomial in $\qq$ and $\rr$ with integer coefficients. Armed with this expression for the 2-pt function, one can carry out the Gram-Schmidt procedure explicitly and obtain the bulk-to-boundary map for 1-particle wormhole states. Alternatively, it is not difficult to compute overlaps for modest $n_\lt, n_\rt$ using the expressions for $T_\lt$ and $T_\rt$ presented in \nref{tl1p}.  

We can compute the boundary states corresponding to chord number eigenstates with $m > 1$ particles in the wormhole $\ket{n_0, \cdots, n_m}_{s_1,s_2, \cdots, s_m}$ using the same ingredients as the above cases. We organize such states into multi-dimensional arrays, following \nref{array_states}. We build the states by induction in $m$, the number of particles. 	Imagine that we know the boundary states corresponding to wormholes with $m-1$ particles. Then by acting on a boundary state with $(M_{s_m})_\rt$, we generate a wormhole with $m$ particles, where $n_m = 0$. Then we essentially follow the same procedure as before to build up states with $n_m > 0$, just like how we built the array \nref{array_states} one column at a time. %

A subtlety for $m>1$ is that states like $\th^l M_s \th^m M_s \th^n\ket{\Omega}$ can generate chord states with 2 matter particles, but also chord states with 0 matter particles. Therefore, one must orthogonalize with respect to states with less matter particles. As simple examples,
\begin{equation}
\begin{split}
\label{}
M_s \th M_s \ket{\Omega} &=  \ket{0,1,0} + \rr \ket{1}  \\
\th M_s M_s M_s  \ket{\Omega} &=   \ket{1,0,0,0} + (2 + e^{-2 s^2/N}) \ket{1,0},
\end{split}
\end{equation}
In the second example, the factor of $e^{-2 s^2/N}$ comes from the $
\wick[wickcolor=dgreen]{
H \c2 M_s M_s \c2 M_s \ket{\Omega} }
$ contraction, see Figure \ref{feynman}.

In summary, we have presented an algorithm for constructing the bulk-to-boundary map that is exact in the double scaling limit. This map is more than just a glorified version of HKLL \cite{Hamilton:2006az} since it works even in the regime where the Schwarzian mode is strongly coupled. We expect the map to be well-defined even beyond the scrambling time. %

\subsection{Measuring the chord number \la{size}}
In this section, we explore the issue of measuring the length of the wormhole, or the total chord number\footnote{For states with matter particles, we could define multiple total chord numbers: the total number of $H$ chords, total number of $M_{s_1}$ chords, total number of $M_{s_2}$ chords, etc. States that differ in any one of these total chord numbers are orthogonal and the comments in this paragraph apply.}.
We showed in Section \ref{bulkBd} that using the explicit bulk-to-boundary map, one can construct explicitly states of any given chord numbers. Furthermore, states with different total chord numbers are orthogonal. So we could reconstruct the operator that measures the total chord number via  $n = \sum_n \ketbra{n}$, where $\ketbra{n}$ is shorthand for a projector onto the subspace of states with a given total chord number. Since we can reconstruct each state in the sum, it follows that we can reconstruct $n$ on the boundary.

While this is formally correct, one might wonder whether there are simpler\footnote{There may also be more complicated ways of measuring the chord number or the geodesic length, see \cite{Stanford:2014jda}.} ways to measure the chord number in the microscopic theory. Note here that we are discussing the double scaled theory, so we are only looking for quantities that track the chord number for a time that does not diverge in the double scaled limit. (Any wormhole effects involving the length mode, like those discussed in \cite{Saad:2019pqd, Iliesiu:2021ari, Stanford:2022fdt} are negligible in this regime).

One idea is to simply use the 2-sided operator $\Psi_{I,\lt}^s \Psi_{I,\rt}^s$. Setting $s = \Delta q$, such an operator  gives\footnote{Unlike measuring the distance using $M_{s,\lt} M_{s,\rt}$, higher powers of the operator \nref{fieldLength} just give $e^{- k \Delta \ell}$. No other Wick contractions appear. I thank David Kolchmeyer for discussing this.}
\begin{equation}
\begin{split}
\label{fieldLength}
{N \choose s}\inv  \sum_I i^s \Psi_{I,\lt}^s \Psi_{I,\rt}^s =  e^{-2s q  \bar{n}/ N} = e^{-\Delta \ell}.%
\end{split}
\end{equation}
On a state with no matter $\bar{n}$ is simply the total number of chords. More generally, acting on a wormhole state with multiple particles, %
\begin{equation}
\begin{split}
\label{normsize}
  \bar{n} \ket{n_0, n_1, \cdots, n_m}_{s_1, \cdots, s_m} = \frac{1}{q} \lp q n_0 + s_1 + qn_1 + s_2 + \cdots+  qn_m \rp \ket{n_0, n_1, \cdots, n_m}_{s_1, \cdots, s_m}\!.
\end{split}
\end{equation}
We believe $\lambda \bar{n}$ is the natural generalization of the length $\ell$ away from the triple scaling limit\footnote{Other options include just $n_H$ or the total number of all chords (without weighting by $s_i$). These operators would also agree with $\tl$ in the triple scaling limit, but based on the form of $T_\lt, T_\rt$ we find these other options implausible.}. 
An interesting feature of the double scaling limit is that there are infinitely many light fields. Therefore, we can take $\Delta $ to be arbitrarily small. Using this method, we can construct a ``length'' operator $(1-\Psi_{I,\lt}^s \Psi_{I,\rt}^s)/\Delta \approx \ell$. Notice however that this approximation only works when $\Delta \ell \ll 1$. By choosing $\Delta \ll 1$ arbitrarily light operators\footnote{Alternatively, we can work at finite $\Delta$ but try to define the logarithm of this 2-sided operator by a ``replica'' trick, e.g., by considering powers $(\Psi_{I,\lt}^s \Psi_{I,\rt}^s)^k$ and then taking $k\to 1$, see Appendix G of \cite{Lin:2019qwu}.}, we can expand the range in which this operator agrees with $\ell$.

This method is morally similar to measuring the distance by looking at correlation functions of some field $\phi$. The downside of this general approach \cite{Lin:2019qwu} is that we are restricted to states where $\phi$ is in the vacuum. If we start perturbing the field $\phi$, our measurements of the length will not be reliable\footnote{This idea is familiar in astronomy. If we measure the distance to some astrophysical object by using its electromagnetic radiation, we must worry about anything that could distort the light along the way. Using multiple types of fields (e.g. neutrinos or gravity waves) gives a more robust measurement.}. Of course, at large $q$ the situation is more favorable because we have a huge number of fields that are arbitrarily light. By summing over these fields, we can create a more robust measurement. In the strict double scaling limit, this measurement becomes perfectly robust if we are restricted to act with operators in the double scaled algebra \nref{sec:dsalg}.

Notice that the formula \nref{fieldLength} was derived in the limit where $\Delta$ is held finite, e.g., $s \propto q$.
We could also wonder whether the formula makes sense when $s= 1$. This is interesting for two reasons. The first is that $s=1$ is the lightest possible operator, so it seems the most favorable for defining the length. Second, the $s=1$ operator in \nref{fieldLength} is the ``operator size''  in the sense of \cite{Qi:2018bje}:
\begin{equation}
\begin{split}
\label{sizeop}
  \text{size} &= \hf \sum_{\alpha=1}^N \lp 1 + i  {\psi}_{\alpha}^\lt \psi_{\alpha}^\rt\rp .%
\end{split}
\end{equation}
Namely, write any 2-sided state $\ket{\chi}$ in the ``size basis'': %
\begin{equation}
\begin{split}
\label{}
  \ket{\chi} = \sum_{s} \sum_{I} c_{s,I} \Psi^s_{I} \ket{\Omega},
\end{split}
\end{equation}
where                                                                                                                                                                                                                                                                                                                                                                                                                                                                                                                                                                                                                                                                                                                                                                                                                                                                                                                                                                                                                                                                                                                                                                                                                                                                                                                                                                                                                                                                                                                                                                                                                                                                                                                                                                                                                                                                                                                                                                                                                                                                                                                                                                                                                                                                                                                                                                                                                                                                                                                                                                                                                                                                                                                                                                                                                                                                                                                                                                                                                                                                                                                                                                                                                                                                                                                                                                                                                                                                                                                                                                                                                                                                                                                                                                                                                                                                                                                                                                                                                                                                                                                                                                                                                                                                                                                                                                                                                                                                                                                                                                                                                                                                                                                                                                                                                                                                                                                                                                                                                                                                                                                                                                                                                                                                                                                                                                                                                                                                                                                                                                                                                                                                                                                                                                                                                                                                                                                                                                                                                                                                                                                                                                                                                                                                                                                                                                                                                                                                                                                                                                                                                                                                                                                                                                                                                                                                                                                                                                                                                                                                                                                                                                                                                                                                                                                                                                                                                                                                                                                                                                                                                                                                                                                                                                                                                                                                                                                                                                                                                                                                                                                                                                                                                                                                                                                                                                                                                                                                                                                                                                                                                                                                                                                                                                                                                                                                                                                                                                                                                                                                                                                                                                                                                                                                                                                                                                                                                                                                                                                                                                                                                                                                                                                                                                                                                                                                                                                                                                                                                                                                                                                                                                                                                                                                                                                                                                                                                                                                                                                                                                                                                                                                                                                                                                                                                                                                                                                                                                                                                                                                                                                                                                                                                                                                                                                                                                                                                                                                                                                                                                                                                                                                                                                                                                                                                                                                                                                                                                                                                                                                                                                                                                                                                                                                                                                                                                                                                                                                                                                                                                                                                                                                                                                                                                                                                                                                                                                                                                                                                                                                                                                                                                                                                                                                                                                                                                                                                                                                                                                                                                                                                                                                                                                                                                                                                                                                                                                                                                                                                                                                                                                                                                                                                                                                                                                                                                                                                                                                                                                                                                                                                                                                                                                                                                                                                                                                                                                                                                          we again use the index $I$ to denote different products of fermions of size $s$. Here $\ket{\Omega}$ denotes the infinite temperature TFD (a maximally entangled state). 
The operator defined in \nref{sizeop} acts on this basis as: $(\text{size})\Psi^s_I\ket{\Omega} = s \Psi^s_I\ket{\Omega}$.

In the double scaling limit, the claim is that
\begin{equation}
\begin{split}
\label{eq:sizechord}
\bar{n} = \frac{1}{q}  \text{size} .%
\end{split}
\end{equation}
A basic property of size is that it has integer eigenvalues in the microscopic theory. Since $\ell = \lambda \bar{n}$, if true, this equation would prove a microscopic explanation for the discreteness of the length!
Now, equation \nref{eq:sizechord} would follow from naively extrapolating \nref{fieldLength} to $s \to 1$ or $\Delta \to 1/q$. 
But since this is outside of the regime of validity in which \nref{fieldLength} was derived, we need to check this claim by reanalyzing the chord diagrams. Let us consider for simplicity the thermofield double state $\ket{\chi} =  e^{-\beta H/2} \ket{\Omega}$. We can compute the size of $e^{-\beta H/2}$ by expanding in powers of $H$. Let's focus on a particular term in the computation of the average size: 
\begin{equation}
\begin{split}
\label{exSize}
& \; \;\;\, \sum_{\alpha=1}^{N} \, \text{tr}\, (
\wick[wickcolor=red]{
\contraction[4ex]{ H H H }{H}{H \psi_\alpha HH}{H}
\contraction[2ex]{ H H  }{H}{H \psi_\alpha H}{H}
\contraction{H H  H  \psi_\alpha}{HHH}{}{H}
\contraction{}{H}{}{H}
 H H H  H \c4 \psi_\alpha H H  H  H \c4  \psi_\alpha })\\
 &\propto \sum_{I, \alpha} \, \wick[wickcolor=red]{\tr \lp  \Psi_{I_1}^{q} \Psi_{I_1}^{q} \Psi_{I_2}^{q}  \Psi_{I_3}^{q}  \c2 \psi_\alpha  \Psi_{I_4}^{q}  \Psi_{I_2}^{q} \Psi_{I_4}^{q}  \Psi_{I_3}^{q} \c2 \psi_\alpha \rp }\\
 &\propto \sum_{I, \alpha} \, \wick[wickcolor=red]{\tr \lp   \Psi_{I_2}^{q}\Psi_{I_3}^q \c2 \psi_\alpha   \Psi_{I_2}^{q} \Psi_{I_3}^q \c2 \psi_\alpha \rp }
\
\end{split}
\end{equation}
Here we have drawn in black a possible chord diagram. We have also drawn a {\color{red} red chord} that corresponds to the (nontrivial part of) the size operator.  After the disorder average, we obtain the product of fermions displayed in the second line. Pairs of fermion subsets which correspond to the chords that do {\it not} intersect the red curve do not contribute to the size, since they simplify via $\psi^2 = 1$. Only the black chords which cross the red chord contributes to the size. Each intersection gives a factor of $q$, the size of the subset of fermions. 
So the net effect of inserting the size operator is to simply multiply each chord diagram by a factor $qn$, where $n$ is the number of intersections between black and red chords (in the example \nref{exSize}, $n=2$). 
Equivalently, we can say that inserting the size operator acts on the chord diagrams by multiplying $\ket{n} \to qn \ket{n}$. Here $\ket{n}$ lives in the Hilbert space defined by slicing open the chord diagram on the red chord. Please compare the red chord with the {\color{red} red curves} in Figure \ref{hilbert}. The endpoints of the particular red curve is determined by where we choose to insert $\psi_\lt$ and $\psi_\rt$.

Actually, there is a small imprecision in the above argument\footnote{I am grateful to Douglas Stanford for pointing this out.}. In the regime where $\lambda \ll 1$, we can say that the operator $\Psi_{I_2}^q \Psi_{I_3}^q$ definitively has size $2q$. But more generally, the probability that $\Psi_{I_2}^q \Psi_{I_3}^q$  has size $2m$ fermions in common (so that the  size is in fact $2q- 2m$) is Poisson distributed
\begin{equation}
\begin{split}
\label{poisson}
  P(m) = \frac{1}{m!} (\lambda/2)^m e^{-\lambda/2}.
\end{split}
\end{equation}
This implies that each chord will contribute to the size by an amount that is somewhat smaller than $q$. The distribution \nref{poisson} has mean $\lambda/2$, so this gives a pair of chords actually contributes size $\sim q - \lambda$. In other words, we imagine that the size distribution of a state $H^n \ket{\Omega}$ to be peaked near sizes $0, q, 2q, \cdots, 2qn$ but each peak has a width of order $\sim \lambda$. (The width of each successive peak grows  $\sim {n \choose 2} \lambda$.) Note however, that if we are interested in the chord number $\bar{n}$, we only care about size$/q$. This means that such effects are $\mathcal{O}(1/q)$, so we neglect them in the strict double scaling limit. 
We should just keep in mind that bulk states with fixed chord numbers are not precisely eigenstates of the size operator, but their size distributions are smeared by a small amount (compared to $q$). Since we checked that the corrections to the entire size distribution are small, this shows that the equation \nref{eq:sizechord} holds as an operator statement. Generalizing to the case with matter particles does not require any important modification of the above arguments. %

\subsection{The double scaled algebra \la{sec:dsalg} }
We can summarize our results in a somewhat more abstract language, which connects to recent papers discussing the emergence of the algebra of bulk fields from the boundary at infinite $N$, see \cite{Leutheusser:2021qhd, Leutheusser:2021frk, Witten:2021unn}. In our context, we can define the Hilbert space of double scaled SYK to be 
\begin{equation}
\begin{split}
\label{dsAlg}
 \mathcal{H} &=  \lim_{\substack{N, \, q \, \to \, \infty \\ \lambda \text{ fixed} }}   \mathcal{H}_\text{SYK} \\
 &= \text{span} \Big\{  H^n \ket{\Omega}, \;\; H^{n_0} M_{s_1} H^{n_1} \ket{\Omega}, \;\; H^{n_0} M_{s_1} H^{n_1} M_{s_2} H^{n_2}  \ket{\Omega}, \;\; \cdots  \Big\}
\end{split}
\end{equation}
One can define an algebra of observables $\hat{\mathcal{A}}$ that is generated by the Hamiltonian and the matter operators $\{ H, M_s, M_{s'}, \cdots \}$. The algebra consists of superpositions of ``words'' of these operators (products of $H$ and $M$ in arbitrary orders) such as $H^{n_0} M_{s_1} H^{n_1} M_{s_2} H^{n_2}$.  In \nref{dsAlg}, $s_i \in \{s, s', \cdots \}$, e.g., we can put any matter operator that we like. 
These operators act on (say the left side of) the infinite temperature TFD state $\ket{\Omega}$, which formally has an infinite entropy $S_\text{max} = N \log 2 \to \infty$ in the double scaled limit. 
Although the entropy is infinite, it is sensible to discuss the entropy deficit $S_\text{max} - S  \propto 1/\lambda$, which is finite.\footnote{In the classical approximation, the low temperature entropy is $S_0$ which differs from $S_\text{max}$ by $\sim 1/\lambda$. So we could equivalently discuss the entropy above extremality $S - S_0$. Note that when quantum corrections are included, the low temperature entropy has logarithmic corrections $\sim \log T$ that make $S_\text{max}- S$ or $S-S_0$ arbitrarily negative. This is just a reflection of the edge of the spectrum $\rho \sim \sqrt{E-E_0}$. Thus in the double scaling limit, the entropy is bounded above $S-S_\text{max} \le 0$ but not bounded below. }

These states are dual to chord states with some number of particles.
More formally, we may define a bulk Hilbert space
\begin{equation}
\begin{split}
\label{dsAlgB}
 \mathcal{H}_\text{bulk}
 &= \text{span} \left\{\ket{n_0}, \;\;  \ket{n_0,n_1}_{s_1}, \;\; \ket{n_0,n_1, n_2}_{s_1,s_2}, \;\; \cdots  \right\}
\end{split}
\end{equation}
The bulk-to-boundary map defined in Section \ref{bulkBd} defines an isomorphism between $\mathcal{H}$ and $\mathcal{H}_\text{bulk}$. Therefore, we can also say that $\mathcal{\hat{A}}$ acts on $\mathcal{H}_\text{bulk}$ by intertwining with the bulk-to-boundary map.

Since $\mathcal{H}$ contains a maximally entangled state with infinite entropy $\ket{\Omega}$, as well as states with smaller (but still infinite) entropies, we expect that the algebra defined above to be of Type II$_1$. (See \cite{Chandrasekaran:2022cip} for a review of Type II$_1$ algebras. In the context of $\qq$-deformed von Neumann algebras, see \cite{Bozejko:1996yv, Ricard, Sniady_2004}.) A Type II$_1$ algebra is equipped with a trace, which we define via 
\begin{equation}
\begin{split}
\label{tr}
  \tr a = \ev{a}{\Omega}, \quad a \in \hat{\mathcal{A}}.
\end{split}
\end{equation}
The fact that the algebra is Type II$_1$ implies that unentangled states (one-sided black holes) are not contained in the double scaled Hilbert space defined above\footnote{It would be interesting to try to define an algebra using the Kourkoulou-Maldacena states \cite{Kourkoulou:2017zaj} instead of $\ket{\Omega}$.}, and that in this limit $\mathcal{H} \ne \mathcal{H}_\lt \otimes \mathcal{H}_\rt$.

We can contrast this double scaled algebra to the large $N$ algebra discussed in \cite{Leutheusser:2021qhd, Leutheusser:2021frk,   Witten:2021unn}.
The analog of the single trace operators in double scaled SYK is the set of operators $\{ M_s, M_{s'}, \cdots \}$. However, a key difference is that {\it the double scaled algebra contains the Hamiltonian}. In $\nn=4$ SYM, the Hamiltonian scales with $N^2$, and therefore is formally infinite and not part of the single trace algebra. In contrast, the Hamiltonian in double scaled SYK scales $\sim N/q^2 \sim 1/\lambda$ which is {\it finite} in the double scaled limit. Note the importance of the second large parameter $q$ which absorbs the large $N$ infinity. 
This would {\it not} be the case for finite $q$ SYK: a straightforward application of the constructions of \cite{Leutheusser:2021qhd, Leutheusser:2021frk,  Witten:2021unn} to finite $q$ SYK at finite temperature would not include the Hamiltonian in the large $N$ algebra and would give a type III algebra \cite{Leutheusser:2021qhd, Leutheusser:2021frk} or a Type II$_\infty$ algebra \cite{Witten:2021unn}. 

Since $H$ is part of the algebra, the thermofield double at any temperature/Lorentzian time $e^{- (\beta/2 + i T) H} \ket{\Omega}$ is included in $\mathcal{H}$, so long as $\beta, T$ are finite in the double scaling limit. The fact that we can discuss both the energy and the entropy above extremality is intimately related to the fact that in this limit, dynamical quantum gravity is still ``turned on,'' albeit in a rather simple form (the Schwarzian mode and its $\qq$-deformation). Note also that for finite $\lambda$ our bulk Hilbert space does not have field-theory divergences due to the lattice regulator that arises from the quantization of the chord number.

In the strict double scaling limit, the rules for computing the trace \nref{tr} are entirely defined by the combinatorics of chord diagrams. In other words, we can completely forget about the microscopic definition of $H$ and $M_s$ in terms of Majorana fermions and random couplings and view the chord diagrams as providing an abstract definition for $\hat{\mathcal{A}}$ and $\mathcal{H}$. 

\subsubsection{The two-sided gravitational subalgebra}
In addition to the matter operators and the Hamiltonian, we may enrich the double scaled algebra $\hat{\ca} \subset \ca $ to include the normalized size/chord number operator $\bar{n}$ defined in \nref{normsize}. This two-sided operator is well-defined in the double scaling limit. The full double scaled algebra $\ca$ then is generated by products of operators
\begin{equation}
\begin{split}
\label{}
 \{  \bar{n},  H_\lt, H_\rt, (M_s)_\lt, (M_s)_\rt, (M_{s'})_\lt, \cdots \}
\end{split}
\end{equation}
Since $\bar{n}$ is a two-sided operator, we should distinguish between operators acting on the left and right, a difference which we glossed over in writing \nref{dsAlg} since $\ket{\Omega}$ is a maximally entangled state.

By considering all possible products of just $\bar{n}, H_\lt$, and $H_\rt$, we can generate a closed subalgebra $\mathcal{G} \subset \mathcal{A}$. It would be interesting to characterize this subalgebra $\mathcal{G}$. In the triple-scaling limit, the results of Section \ref{multiparticles} show that $\mathcal{G}$ is precisely the gravitational algebra of JT gravity, including an \slt\, algebra, where the renormalized length $\tilde{\ell} = \lambda \bar n + 2 \log \lambda$. We checked the gravitational algebra \nref{harlow_wu} for states of arbitrary numbers of matter particles, which span the entire Hilbert space \nref{dsAlg}. Therefore \nref{harlow_wu} must hold as an operator statement on $\mathcal{G}$. Presumably away from the triple-scaling limit, the gravitational algebra is $\qq$-deformed, and correspondingly the physical \slt\, algebra becomes an \slt$_\qq$ algebra\footnote{As evidence of this, note that the $T$ in \nref{transferBerkooz} has the form of a $\qq$-deformed Liouville Hamiltonian \cite{Olshanetsky:1993sw}.}. 
The existence of a physical \slt \, subalgebra is a smoking gun signature of an emergent AdS$_2$ geometry, so understanding this $\qq$-deformation is important for understanding the bulk dual away from the double scaling limit. We will report on this in a future paper. %

Note that in \cite{Leutheusser:2021qhd,Leutheusser:2021frk} the near horizon symmetries of the black hole background were given an algebraic construction using the half-sided modular inclusion. Here also we have constructed the near horizon symmetries; it would be nice to relate these constructions.

\section{Discussion}
We conclude with some further comments and future directions.
\begin{enumerate}
	\item {\it Tensor networks}: the chord diagram is the more precise version of the idea that there is some sort of tensor network in the bulk. In AdS$_2$/CFT$_1$, 2-sided states should be described by a 1D tensor network. A 1D tensor network is just a product of matrices that represents the 2-sided state. Here one can say that the chord diagram is also a graphical representation of a product of matrices, with the number of chords recording the length of the product. The ``dangling legs'' of the tensor network are simply the choices of what particle type to insert. Note here that our formulas give rules for ``updating'' the tensor network after time evolution.
	\item[1a.] Along these lines, tensor networks are meant to be a toy model of the bulk-to-boundary map that illustrate certain information theoretic properties. But now that we have concrete formulas for the bulk-to-boundary map, it would be nice to directly characterize its error-correcting properties  \cite{Almheiri:2014lwa, Chandrasekaran:2022qmq}.
	\item The chord diagrams in the JT limit seem to prefer geodesic slices of the bulk. How do we describe other bulk slices within the same Wheeler de Witt patch? 
	\item {\it Entropy}: the dilaton is not manifest in our discussion of the chord diagrams. How does it emerge from the chords? A related question is whether there is a QES/JLMS formula that directly involves the chord diagrams. 
	\item {\it Wormholes}: In the $N\to \infty$ limit, wormhole effects are negligible. Nevertheless, in \cite{Baur} the authors computed $\qq$-deformed correlators of probe fields on JT wormholes; it would be nice to reproduce them using the chord combinatorics.
	\item {\it SYK variants}: It would be interesting to study various generalizations of the double scaled SYK model, especially the $\nn =2$ version \cite{Berkooz:2020xne}. Presumably the ground state sector of the $\nn=2$ model would lead to a Type II$_1$ subalgebra of the full double scaled Type II$_1$ algebra $\mathcal{A}$. This might shed light on the emergence of bulk time \cite{LongPaper, Lin:2022rzw}.
	\item {\it Cosmology}: Understanding the emergence of the bulk Hilbert space in cosmology is an outstanding problem. It was conjectured in \cite{Susskind:2021esx, VerlindeDS} that double scaled SYK has a de Sitter interpretation\footnote{According to \cite{Susskind:2021esx}, we should see the de Sitter picture emerge at times of order $q$ where an extrapolation of the chord results has not been justified.}. Our results show that the double scaled model has a Type II$_1$ algebra, in line with de Sitter expectations \cite{Chandrasekaran:2022cip, Lin:2022nss}.  On the other hand, the chord picture does not obviously resemble a de Sitter geometry. This should be explored.

\end{enumerate}

\subsection*{Acknowledgments}
I thank Ahmed Almheiri, Micha Berkooz, Daniel Harlow, Adam Levine, Stephen Shenker, Edward Witten, and Zhenbin Yang for stimulating discussions.

I especially thank Juan Maldacena for comments on the draft, David Kolchmeyer for tutoring me on $\qq$ deformations, and Douglas Stanford for bringing Krylov complexity to my attention. I thank both Douglas and David for seriously helpful discussions about the inner product of chord states. I thank Douglas Stanford, Cynthia Yan, and Edward Witten for pointing out typos/errors in previous versions of the paper.
I am supported financially by a Bloch fellowship.

\appendix

\section{Classical limit}\la{Classy}
Here we consider the classical limit of the Hamiltonian given in \nref{largeqHam}.
This is a limit where $\lambda k $ is held fixed, but $\lambda \to 0$ (equivalently $\qq \to 1$). This gives: 
\begin{equation}
\begin{split}
\label{eq:classy}
  H \approx \frac{2}{\lambda }\sqrt{1- e^{-\ell}}\cos (\lambda k).
\end{split}
\end{equation}
This Hamiltonian was derived in \cite{Lensky:2020fqf} by analyzing the classical equations of motion in the large $q$ SYK model. In their analysis $\ell$ is propertional to $g$, which determines the two point function in SYK $G \propto e^{g/q}$. %
 Here we see that the appropriate quantum generalization is \nref{largeqHam}.  %
Actually, \cite{Lensky:2020fqf} derived \nref{eq:classy} up to a factor of $2q^2/N$ which is classically ambiguous.
(Hamilton's equations for $H = 1/a K(a p)V(\ell)$ give $\dot{\ell} =  k'(ap) V(\ell), a\dot{p} =  -k(ap) V'(\ell)$. By redefining $\tilde{p} = a p$ we get the same equations of motion as for $a=1$.) The quantum generalization sets the period of the cosine, which in turn tells us about the discreteness of the $\ell$ variable. 

A peculiarity of this Hamiltonian is that despite the complicated-looking form, the equations of motion are still Liouville-like.
Applying Hamilton's equations,
\begin{equation}
\begin{split}
\label{}
  \ddot{\ell} = \pd_\ell \pd_k H \pd_k H - \pd_k^2 H \pd_\ell H = 2 e^{-\ell}.
\end{split}
\end{equation}
This agrees with the fact that at finite energies, the 2-pt function in large $q$ SYK satisfies a Schwinger-Dyson equation that has the Liouville form \cite{Maldacena:2016hyu}. 

\section{Derivation of the quantum gravitational algebra \la{app:quantumwu}}
In this appendix we derive the gravitational algebra of Harlow and Wu \cite{Harlow:2021dfp} using the particle-in-a-group formalism \cite{Yang:2018gdb, Kitaev:2018wpr}. The advantage of this derivation is that it holds in the quantum Schwarzian regime with arbitrary quantum matter in the bulk. 

We will use the conventions of \cite{LongPaper} and set $C=1/2$, see Section 4.1 of \cite{LongPaper} for a review. The basic idea is that one views the boundary particle as a particle on the group coset \slt$/U(1)$, which is AdS$_2$. The propagator on this coset defines the action of $e^{-\beta H}$, or equivalently the two-sided thermofield double.
If we write $g= e^{-x L_-} e^{\rho L_0} e^{\gamma L_+}$ the Casimir is given by
\begin{equation}
\begin{split}
\label{}
\mathcal{C}=-\mathcal{L}_{0}^{2}+\frac{1}{2}\left(\mathcal{L}_{+} \mathcal{L}_{-}+\mathcal{L}_{-} \mathcal{L}_{+}\right)=\partial_{\rho}^{2}+\partial_{\rho}+e^{-\rho} \partial_{x} \partial_\gamma.
\end{split}
\end{equation}
We consider two Schwarzian particles representing the two sides of the wormhole, doubling the number of coordinates.
To account for the $U(1)$ quotient, we gauge the $U(1)$ generator by setting $\pd_{\gamma_1} = +1$ and $\pd_{\gamma_2} = -1$. Hence we have the expressions
\begin{equation}
\begin{split}
\label{}
  H_\lt = -\mathcal{C}_1 - \frac{1}{4} = -  \partial_{\rho_1}^{2}-\partial_{\rho_1}-e^{-\rho_1} \partial_{x_1} - \frac{1}{4}\\
  H_\rt = -\mathcal{C}_2 - \frac{1}{4} = -  \partial_{\rho_2}^{2}-\partial_{\rho_2}+e^{-\rho_2} \partial_{x_2} - \frac{1}{4}\\
\end{split}
\end{equation}
The renormalized length between the two points is
\begin{equation}
\begin{split}
\label{}
{\tl} = {\rho_1 + \rho_2} + 2\log (x_1-x_2)
\end{split}
\end{equation}
Furthermore, we can define
\begin{equation}
\begin{split}
\label{}
- 2 i  k_\lt &=  [H_\lt, \tl] = 1+ \frac{2 e^{-\rho_1}}{x_1-x_2} + 2 \pd_{\rho_1}\\
- 2 i  k_\rt &=  [H_\rt, \tl] = 1+ \frac{2 e^{-\rho_2}}{x_1-x_2} + 2 \pd_{\rho_2}.
\end{split}
\end{equation}
With these definitions, one can check \nref{harlow_wu} with $C=1/2$.
Notice that in this derivation, we did not use the gauge constraint on $(\mathcal{L}_m)_1 + (\mathcal{L}_m)_2$. Hence the derivation holds for an arbitrary amount of matter.

As an aside, this method should allow us to derive the gravitational algebra in cases with more supersymmetry, where the JT action is cumbersome.%

\pagebreak

\bibliographystyle{apsrev4-1long}
\bibliography{bibChords.bib}
\end{document}